\documentclass[final]{aa}  

\usepackage{graphicx}
\usepackage{txfonts}
\usepackage{subcaption}         
\usepackage{lscape}             
\usepackage{booktabs}
\usepackage{stfloats}
\usepackage{placeins}           
                                
\usepackage[colorlinks=true,allcolors=blue]{hyperref}

\begin{document}
\nolinenumbers
   \title{Optimising instrument concepts with Machine Learning: Application to CMB spectral distortion experiments}


%
%
%

   \author{X. Coulon\inst{1,2}\corrauth{xavier.coulon@universite-paris-saclay.fr}        
        \and N. Aghanim\inst{1}
        \and B. Maffei\inst{1}
        }

   \institute{Université Paris-Saclay, CNRS, Institut d'Astrophysique Spatiale, 91405 Orsay, France
   \and Centre national d’études spatiales (CNES), France}

   \date{Received September XX, 20XX}

 
  \abstract
    {Understanding how instrumental and mission parameters affect the ability to measure faint astrophysical signals is a key challenge in the design of future experiments. This is particularly relevant for CMB spectral distortions, whose weak signals are affected by both instrumental effects and astrophysical foregrounds. We develop a method to explore and optimise the multidimensional parameter space of an astronomical instrument, and apply it to CMB spectral distortion measurements using the FOSSIL mission concept. We combine a dedicated sky model, including CMB spectral distortions with both galactic and extragalactic foregrounds, with a realistic spectro-photometric instrument model and use Fisher forecasts to assess measurement capabilities of the simulated instrument. Decision-tree-based regression models are then used to learn the non-linear mapping between instrumental parameters and the predicted signal-to-noise ratios of the sky observables, in particular CMB spectral distortions and the Cosmic Infrared  Background. Random Forest and gradient boosting models accurately reproduce the forecasted measurement performance. SHAP values are used to interpret the impact of instrumental parameters on the measurement. The temperature of the warmest instrumental component is found to be the dominant parameter for all three observables, highlighting the importance of limiting internal emission and operating the instrument at cryogenic temperatures. Frequency coverage is also highly influential, revealing a trade-off between spectral coverage and instrumental sensitivity.The proposed optimisation method provides a fast and interpretable approach to explore high-dimensional instrumental parameter spaces and allowing for the identification of the parameters that most strongly influence the scientific performance of an experiment. To the best of our knowledge, this work represents the first application of machine-learning methods to the global optimisation of an astronomical instrument and can readily be extended to other astronomical instruments and mission concepts.}
    
    

   \maketitle


\nolinenumbers

\section{Introduction}
\label{sec-0:intro}

Understanding how instrumental model parameters affect the predicted ability to measure astrophysical signals is essential for the design and performance evaluation of future experiments \citep{Tauber2010planck_model_instru, Laureijs2011euclid_model_instru,Braun2019SKA_tech_report}. The importance of such an approach is even greater when the targetted signals are intrinsically low (e.g., gravitational waves \cite{Amaro2017LISA}) or if they are limited by contaminations from astrophysical sources or systematic effects (e.g., Cosmic Microwave Background (CMB) anisotropies \cite{Planck2014HFIbeam_model_lim,Planck2014HFIproc_model_lim, Planck2016X_compo_sep}).\\
These signals from the tiny departures from the CMB blackbody spectrum, also called CMB spectral distortions (see \citet{chluba2021new_horizons} et \citet{Cyr2026SciencePaper} for reviews), suffer from those two aspects: They represent a very low signal and are limited by both the astrophysics and instrumental effects. As a result, missions aiming to detect or constrain the monopole amplitudes of the $y$- and $\mu$-type distortions (see Fig.\ref{fig:SEDs}), require quantifying as best as possible how variations in the spectro-photometric model of the instrument propagate into forecasts for key sky observables. \\
Two main strategies have been widely adopted in the community to quantify parameter dependencies and thereby optimise instrument concepts. On the one hand, based on analytic computations, a systematic impact analysis of instrument parameters on the measurement provides a principled way of identifying which components of the instrumental description exert the strongest influence on the predicted measurement capabilities (e.g., the overall sensitivity of an instrument \citep{Prince2002_opti_LISA} or detector characteristics \citep{Dorigo2023_opti_AD_vs_ML}). 
On the other hand, based on numerical computations, a brute-force approach consists in sampling the instrumental parameter space, computing corresponding forecasts for the astrophysical observables, and evaluating the multi-dimensional correlation structure among parameters (e.g., statistical-based optimisation of telescope layout \cite{CTA2019_opti_CTA} or  sampling based optimisation of detectors \cite{Arectout2021_opti_det, Knapp2023_opti_plasma}). \\
Both approaches are hampered by the increasingly large number of parameters (typically of the order of a few tens for simple models, and easily exceeding hundreds parameters for more realistic descriptions) 
intervening either in the sky of in the instrument models. In the first approach, this calls for exact analytic modelling of the correlations between parameters. The second, and conceptually simple, approach is computationally demanding and sensitive to the choice of parameter intervals, making its rigorous application challenging in practice.\\
An alternative strategy to address the interplay between instrument models and measurement is based on Machine Learning (ML). ML is nowadays commonly used in astrophysical data analysis to detect and classify objects \citep{Bonjean2020_ML_SZ_detection_exemple,Farid2023_ML_galaxy_cluster_classification_exemple, Bonnaire2020_ML_filament_detection_exemple}, to accelerate and enhance numerical simulations \citep{He2019_ML_simulation_acceleration_exemple,Wadekar2022_ML_simulation_emulator_exemple,Conceicao2024_ML_cosmological_emulator_exemple}, or to emulate and generate data \citep{Sethuram2023_ML_radiative_transfer_emulator_exemple,Mikuni2022_ML_generative_detector_simulation_exemple,Alsing2019_ML_simulation_based_inference_exemple,Cranmer2020_ML_simulation_based_inference_exemple}. By efficiently exploring high-dimensional parameter spaces, ML methods can reveal non-linear dependencies. This property can thus help identify optimal instrumental configurations. Therefore, in addition to these standard usages for data analysis, ML has become an increasingly powerful tool for the design and optimization of complex devices, from electronic circuits and integrated chips to large-scale engineering systems \citep{Mirhoseini2021_ML_opti_chip_design_exemple,Baydin2021_ML_opti_experimental_design_exemple,GranadosOrtiz2021_ML_opti_design_optimization_exemple,Qasim2024_ML_opti_physics_instrument_design_exemple}. 
In recent years, these techniques have also been applied successfully to the optimization of telescope subsystems, such as adaptive optics systems \citep{Landman2020_ML_opti_adaptive_optics_exemple,Pou2022_ML_opti_adaptive_optics_exemple}, and to the design and performance optimization of detector subsystems \citep{Qasim2024_ML_opti_detector_design_exemple}, where the interplay between instrumental parameters and scientific observables is often intricate and computationally costly to assess.\\
The optimisation of a Fourier Transform Spectrometer (FTS) for CMB spectral distortion measurements is a high-dimensional problem, where instrumental, mission, and sky-model parameters are strongly interdependent and often degenerate, preventing a purely intuitive or fully analytical assessment of the overall scientific performance. In this work, we present an unsupervised ML-assisted optimisation designed to explore this complex parameter space and identify the instrumental and mission parameters that drive the performance of an FTS-based experiment dedicated to CMB spectral distortion measurements, as considered for missions such as PIXIE \citep{kogut2012pixie}, PRISTINE \citep{pristine_ias}, BISOU \citep{maffei2021bisou} or FOSSIL \citep{aghanim2022fossil, Aghanim2026FOSSIL_general_paper}. More specifically, we focus on the FOSSIL space mission concept proposed as an ESA M8 mission \citep{Aghanim2026FOSSIL_general_paper}, designed to measure the $\mu$ and $y$ CMB spectral distortions while simultaneously constraining the Cosmic Infrared Background (CIB) \citep{Puget1996cib_COBE,Lagache2003cib,Dole2004cib}, together with Galactic foregrounds \citep{Bennett2003_sync_wmap,Draine2003_IDG,Planck2016X_compo_sep} and extragalactic foregrounds \citep{DeZotti2010_radio_sources,Planck2014XII_diffuse_compo_sep}.
Combining a physically motivated sky model, detailed in \cite{Coulon2026FOSSIL_SKY} and briefly described in Sect.~\ref{sec:1-sky}, with a simplified yet realistic spectro-photometric model of the instrument and its main design parameters (Sect.~\ref{sec:instrument_concept}), we derive noise and sensitivity models (Sect.~\ref{sec:sensitivity_model}) and predict the measurement capabilities of FOSSIL through Fisher forecasts \cite[see][for more details]{Coulon2026FOSSIL_SKY}.
The remainder of this paper is devoted to the optimisation of a FOSSIL-like instrument for the measurement of the monopole of CMB spectral distortions and the CIB. In Sect.~\ref{sec:one-param}, we illustrate an example of a one-parameter-at-a-time optimisation strategy together with its limitations. In Sect.~\ref{sec:ML-param}, we introduce a machine-learning-based global optimisation analysis, using decision-tree regression models to quantify how instrumental and mission parameters propagate into Signal-to-Noise Ratio (SNR) forecasts for CMB spectral distortion and CIB observables, thereby overcoming the limitations of brute-force parameter-space exploration. In Sect.~\ref{sec:app_case}, we show how this framework allow us to identifies the parameters that most strongly influence the scientific performance of FOSSIL and highlights the regions of the multidimensional design space that maximise the mission's scientific return. To the best of our knowledge, this is the first application of such a machine-learning-based global sensitivity analysis framework to guide the optimisation of an astrophysical instrument design. Finally, we discuss the proposed methodology in Sect.~\ref{sec: discu} and present our conclusions in Sect.~\ref{subsec: conclu}.


\section{Millimetre and sub-mm sky monopole model}
\label{sec:1-sky}

We adopt a parametric sky model of the microwave and sub-mm sky designed to consistently describe both CMB spectral distortions and the dominant astrophysical foregrounds as in \cite{abitbol_prospects_2017}. A comprehensive description of the model is provided in \cite{Coulon2026FOSSIL_SKY}. It includes implementation aspects, notably the inclusion of spatially varying foregrounds, and a description of the forecasting methodology. 

\begin{figure}[!ht]
    \centering
    \includegraphics[width=\linewidth]{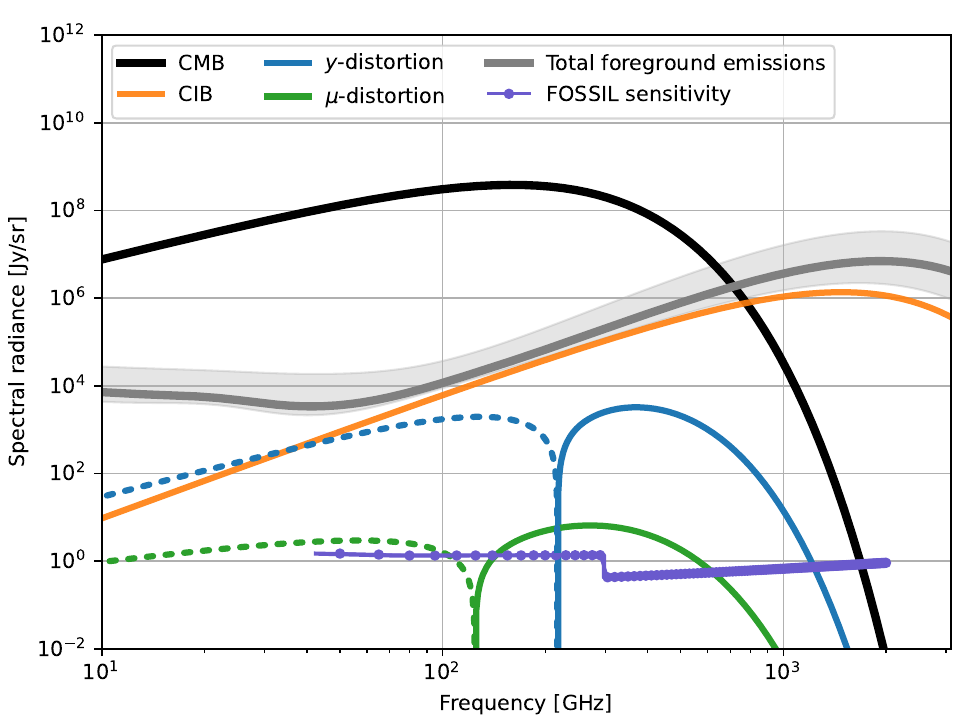}
    \caption{Absolute value of the emission spectra of the signal of interest in this study in the reference sky model used in this work: blackbody emission of the CMB (black), $y$-type distortion monopole (blue), $\mu$-type distortion monopole (green)  and the CIB monopole (orange). The total foreground signal (solid grey line) corresponds to the sum of all foreground emissions averaged over 70\% of the sky, corresponding to the fraction not masked by the Galactic plane. The shaded region indicates the envelope defined by the mean total sky emission over the full sky (upper limit) and 20\% of the cleanest sky (lower limits). The FOSSIL sensitivity is also shown for reference }
    \label{fig:SEDs}
\end{figure}

The sky model used throughout the paper is such that the sky-averaged intensity monopole is defined, after subtraction of a reference blackbody spectrum at temperature $T_0 = 2.725\,\mathrm{K}$ \citep{fixsen2009cmbSD}. It is written as
\begin{align}
    I_\nu = \Delta B_\nu + \Delta I_\nu^{\mu} + \Delta I_\nu^{y} + \Delta I_\nu^{\mathrm{rel\text{-}tSZ}} + I_\nu^{\mathrm{fg}} \, ,
\end{align}
where the different terms correspond to a blackbody temperature shift ($\Delta B_\nu$), the $\mu$-type distortions ($\Delta I_\nu^{\mu}$), the $y$-type distortions  ($\Delta I_\nu^{y}$), relativistic corrections to the thermal Sunyaev--Zel'dovich effect ($I_\nu^{\mathrm{rel\text{-}tSZ}}$), and the emissions of astrophysical foregrounds ($I_\nu^{\mathrm{fg}}$). The spectral radiance of the sky components of interest for this paper is shown in Fig.~\ref{fig:SEDs}.
\\
Galactic contributions are based on the Python Sky Model (PySM) \citep{Thorne2017pysm,Zonca2021pysm3} and include thermal dust emission \citep{Draine2011FreeFree} modelled as a grey body, synchrotron emission \citep{Haslam2001_Galactic_synchrotron_foreground, Bennett2003_sync_wmap, Planck2020_CMB_foreground_components}), free-free emission \citep{Dickinson2003ff_template, Draine2011FreeFree} both modelled as power laws, and Anomalous Microwave Emission (AME)  \citep{Kogut1996ame_fir, Oliveira1998, Lagache2003, Battistelli2019, Arce2020}. Extragalactic emissions include the CIB (orange), modelled as in \cite{abitbol_prospects_2017}, and the cumulative emission from extragalactic CO lines \citep{Wright1991co_mw,fixsen1998lim_firas}, modelled as in \cite{Mashian2016_CMB_CO}. Zodiacal thermal and scattered emissions \citep{Werner2004_Spitzer,Krick2012_ZodiacalIRAC,Planck2014zodi} can also be modelled based on \citep{Planck2011XXame_new}, but are not considered in the remainder of the work.

This formulation provides a unified description of cosmological and astrophysical contributions relevant for spectral distortion monopole measurements \citep{chluba2017moment_expansion}. In contrast to the commonly use approach in CMB  spectral distortion studies based on sky-averaged quantities over a given fractions of the sky \citep{abitbol_prospects_2017}, we explicitly account for spatial variations of Galactic foregrounds using PySM-based templates \citep{PySM3,Thorne2017pysm, Zonca2021pysm3}. This enables a more realistic estimation of the instrument noise limit and treatment of observational strategies for the mission optimisation.

The detailed physical modelling of each component is fully described in \cite{Coulon2026FOSSIL_SKY} together with all the fiducial sky parameters and reference values. In the present study, we focus on the sky parameters listed in Table~\ref{tab:fiducial_values_sds_foregrounds} and investigate the impact of the instrument on their detectability.

%
%
\begin{table*}[!ht]
\centering
\caption{Fiducial values adopted for CMB spectral distortions and CIB.}
\label{tab:fiducial_values_sds_foregrounds}
\renewcommand{\arraystretch}{1.2}
\setlength{\tabcolsep}{8pt}
\begin{tabular}{l l l c}
\toprule
\textbf{Emission}  & \textbf{Model}  & \textbf{Parameter} & \textbf{Value} \\
\midrule\midrule
$y$-type distortion & moment expansion& $y$ & $1.77 \times 10^{-6}$ \\
$\mu$-type distortion & moment expansion& $\mu$ & $2 \times 10^{-8}$ \\
\midrule
Cosmic infrared background & modified blackbody & $A_{\mathrm{CIB}}$ & $3.46 \times 10^{5}~\mathrm{Jy\,sr^{-1}}$ \\
& & $\beta_{\mathrm{CIB}}$ & $0.86$ \\
& & $T_{\mathrm{CIB}}$ & $18.8~\mathrm{K}$ \\
\bottomrule
\end{tabular}
\end{table*}


\section{Instrument concept and model}
\label{sec:instrument_concept}

In contrast with differential measurements of anisotropies performed by imaging missions such as \textit{Planck} \citep{planck2014overview} or \textit{LiteBIRD} \citep{LiteBIRD2023_PTEP}, measurement of CMB spectral distortions requires an absolute determination of the sky intensity as a function of frequency.
%
%
The CMB spectral distortion measurement (in particular the measurement of the $\mu$-distortion monopole) relies on an instrument capable of performing high-precision absolute spectroscopy over a wide frequency range (GHz to THz), with tens to hundred spectral channels, high sensitivities of the order of 1~Jy/sr or below, and a few degrees-scale angular resolution.

The measurement of CMB spectral distortions, in particular the $\mu$-distortion monopole, requires an instrument capable of performing high-precision absolute spectroscopy over a broad frequency range, from GHz to THz, with tens to hundreds of spectral channels, sensitivities of order $1$~Jy/sr or better, and an angular resolution of a few degrees.In this context, a wide-band FTS in a Martin--Puplett configuration \citep{Martin1970_martin_puplett} provides a particularly well-suited solution and efficient implementation of the above-mentioned requirements.
This concept, originally used for \textit{COBE}/\textit{FIRAS} \citep{Mather1990_COBE_FIRAS}, subsequently developed in \textit{PIXIE} \citep{kogut2012pixie,kogut2016pixie,Kogut2024PIXIE}, is adopted in several proposed experiments such as BISOU \citep{Maffei2024_BISOU_pathfinder}, COSMO \citep{Masi2021_COSMO}, FOSSIL \citep{Aghanim2026FOSSIL_general_paper}).

\begin{figure}[!ht]
    \centering
    \includegraphics[height=11cm]{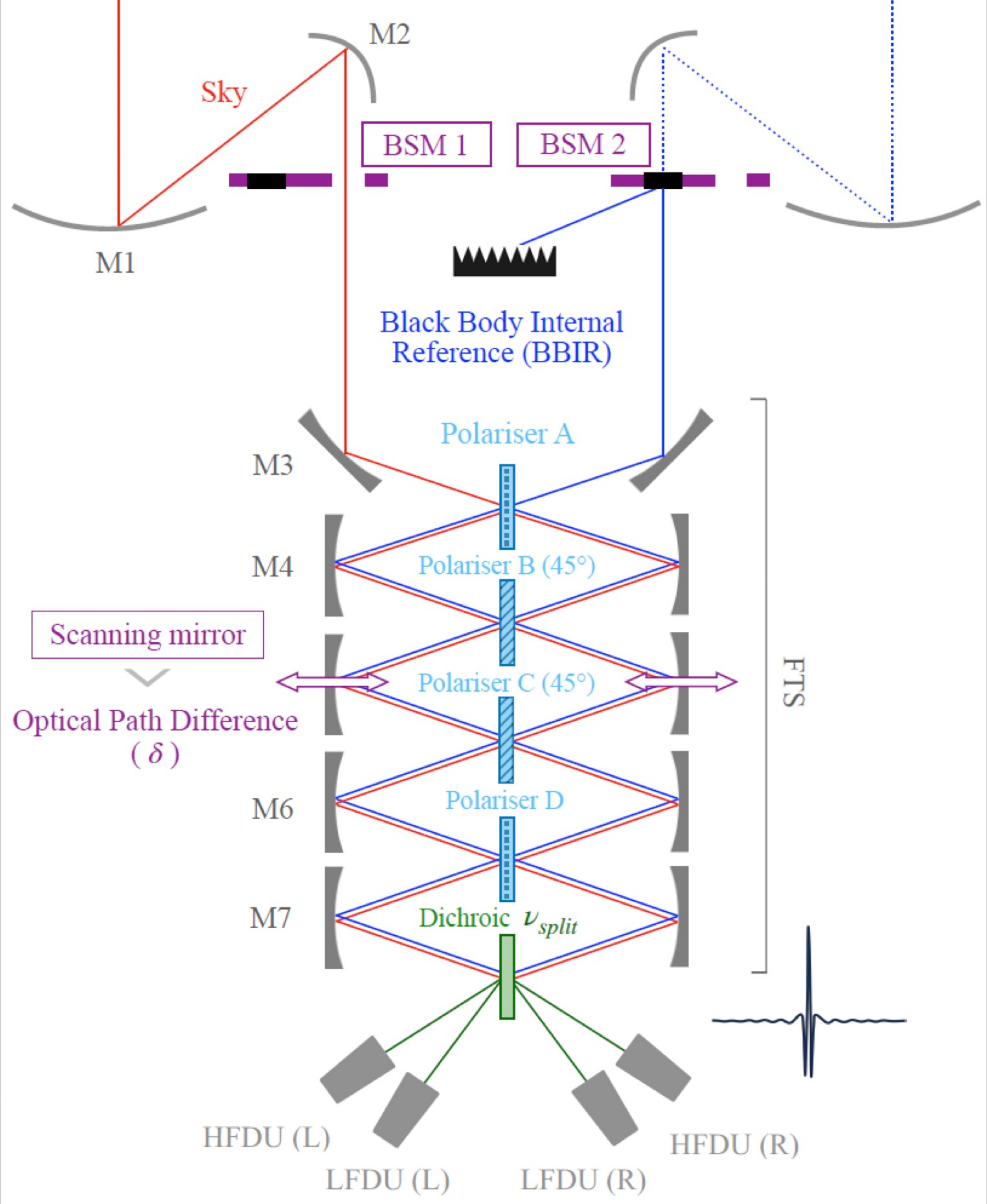}
    \caption{Each telescope consists of a primary mirror ($M_1$), and secondary mirror ($M_2$). The Fourier Transform Spectrometer (FTS) is composed of four polarisers ($P_A$ to $P_D$) and five mirror pairs ($M_3$ to $M_7$), one of which includes the scanning mirror. One of the FTS input is directed towards the sky through the two-mirror telescope, while the second input is directed towards the absolute blackbody reference. The displacement of the scanning pair of mirror induces an optical path difference between the right and left arms of the instrument, enabling interference between the two outputs. Adapted form \cite{LeGall2026optics_BISOU_breadboard}.}
    \label{fig: FTS}
\end{figure}

The instrument considered in this work, a wide-band Martin--Puplett FTS, corresponds to the proposed concept for the FOSSIL mission. The measurement principle follows the approach pioneered by \textit{FIRAS} \citep{Mather1993firas}: one input port of the FTS is pointed towards the sky, while the second points to a reference blackbody source whose temperature is close to that of the CMB. 
The incoming radiation is split into orthogonal linear polarisations by input polarisers, recombined after propagation through two arms with a variable optical path difference, and finally directed towards two output ports equipped with detectors (see Fig.~\ref{fig: FTS}). As the moving mirror introduces an optical path difference, each detector records an interferogram (IFG) as a function of this delay. The Fourier transform of this signal yields the frequency spectrum of the difference between the sky and the calibrator. The instrument therefore measures the difference between the sky emission and the absolute reference, enabling an absolute determination of the spectrum.

The observing direction, $\vec{p}$, together with the beamwidth determines the observed sky region and are not treated as free parameters. The scanning strategy is often fixed by the scientific objectives of the mission and the observing platform (ground-based, balloon-borne, or space-based). Throughout this work, these parameters are thus excluded from further optimisation and we adopt observing strategy described in \cite{Coulon2026FOSSIL_SKY}.
\\

In the following, we present a flexible parametrisation of a spectro-photometric model of the FOSSIL-like instrument for further performance optimisation. To do so, the instrument is decomposed into a set of functional subsystems, each associated with specific physical constraints and performance drivers, and described by a set of parameters (see Table. \ref{tab:fiducial_values_sds_foregrounds}). This decomposition enables the translation of high-level scientific requirements into a set of instrumental parameters. The main subsystems involved in a FOSSIL-like instrument are described in \cite{Aghanim2026FOSSIL_general_paper}.

\subsection{FTS}
\label{subsubsec:fts_mtm}

The central element of the instrument is the FTS which includes the optical assembly (five mirror pairs and four polarisers) and the mechanical system associated with the moving mirror hence the scanning mechanism.
\\
The physical size of the optical elements, in particular the mirrors, is constrained by diffraction, which applies to all optical components of the instrument. Consequently, the primary-mirror size, $D_{\rm{prim}}$, defines the beam throughput and thus the trade-off between sensitivity, angular resolution, and instrument size. 

The FTS scanning mechanism determines the spectral performance of the FTS. Its displacement defines two fundamental instrumental parameters. The first parameter is the maximum optical path difference, $\delta_{\mathrm{max}}$, which directly sets the spectral resolution of the instrument, $\Delta \nu$. A longer scanning stroke increases the maximum optical path difference (OPD) of the FTS  and therefore improves the spectral resolution,  according to:
\begin{equation}
\Delta\nu = \frac{1}{2\,\mathrm{OPD}_{\max}}
\end{equation}
, where $\mathrm{OPD}_{\max}$ is the maximum optical path difference covered by the FTS scan. In practice, FTS instruments can operate in multiple scan modes, as demonstrated with COBE/FIRAS \citep{Stark1986_MTM,Mather1993FIRAS_MTM}, where long and short scans provided different resolutions depending on the frequency range. The second parameter is the sampling step of the mirror displacement, $\Delta \delta$ that determines the highest observable frequency through the Nyquist criterion:
\begin{equation}
\nu_{\mathrm{max}} \leq \frac{1}{2\,\Delta \delta}.
\end{equation}
This constraint arises from the discrete sampling of the IFG, which must be adequately reconstructed in Fourier space. In practice, the sampling of the scanning mirror position is chosen such that this criterion is satisfied with a comfortable margin.
\\ 
Finally, the speed of the scanning mechanism, which is closely linked to the detector acquisition rate, affects the total acquisition time and therefore the effective sensitivity of the instrument.

\subsection{Absolute Reference}
\label{subsubsec:calibration}

Unlike CMB imagers, which measure relative fluctuations, spectral distortion measurements require an absolute reference of known brightness. This is achieved through a blackbody reference with a well-characterised temperature $T_{\mathrm{AR}}$. The instrument then measures the difference between the sky signal and this absolute reference, effectively converting the measurement into a differential comparison between two blackbodies. In both the FOSSIL and BISOU instrument concepts, the blackbody reference is internal, integrated within the instrument and observed via one of the FTS inputs. In the FOSSIL design, however, the internal reference can be viewed through either FTS input using a beam-switching mechanism. In the PIXIE design, the blackbody reference is external \citep{kogut2012pixie}. 

The uncertainty on the absolute reference temperature, $\Delta T_{\mathrm{AR}}$, contributes to the uncertainty on the recovered spectral distortions, since these are inferred from the comparison between the sky signal and the reference blackbody. Consequently, the performance of the absolute reference subsystem is a key driver of the overall instrument performance.
\subsection{Optical components}
\label{subsubsec:optics}

In addition to the mirrors and polarisers of the FTS, spectral filters are needed to define the observational band and to prevent out-of-band radiation from contaminating the signal, particularly near zero optical path difference where all frequencies contribute coherently. Moreover, optical dichroics can be placed after the telescope or before the detection units to separate spectral channels. These elements improve the overall instrument sensitivity (see Sect.~\ref{sec:sensitivity_model}). 
\\
At millimetre and sub-millimetre wavelengths, thermal emission from optical components becomes a dominant source of contamination. To mitigate this effect, the instrument is cooled to cryogenic temperatures, typically of the order of a few kelvin \citep{Sauvage2026FOSSIL_thermal}. A temperature of $\sim 3$~K reduces both the amplitude of thermal emission and shifts its spectral peak towards lower frequencies, thereby reducing contamination in the CMB observation band. Focal plane assemblies are further cooled to sub-Kelvin temperatures ($\sim100$~mK) to ensure photon-noise-limited detector performance.
\\

Each optical element, $k$, is modelled as a grey body characterised by its temperature $T_k$ and emissivity $\epsilon_k(\nu)$. It is also  characterised by a frequency-dependent transmission, $t_k(\nu)$. The overall instrument response is thus described by the product of the individual transmissions. The resulting bandpass is approximated by a trapezoidal profile, featuring a flat response across the nominal frequency range and linear transitions at the band edges. This simplified model provides a realistic first approximation  of the instrument spectral response and will be refined as more realistic simulations and actual measurements of the optical components become available.

\subsection{Focal plane and detectors}
\label{subsubsec:fpa}

The two output ports of the FTS are directed onto a focal plane assembly (FPA) made of detector units, which host arrays of detectors operating at sub-Kelvin temperatures and coupled to multimode feedhorns (see \cite{Aghanim2026FOSSIL_general_paper}). 
\\
Multimode horns were successfully used in \textit{Planck} \citep{Murphy2010_multimode_planck} and have been selected for the COSMO mission concept \citep{Manzan2024_CMBspecDist_COSMO} to maximise the collected power per detector. They are particularly well suited to CMB spectral distortion experiments, for which angular resolution is not a primary science driver since multimode coupling increases the optical throughput, $A\Omega$, thereby improving the instrument sensitivity at the expense of angular resolution. The number of propagating modes, $N$, will increases with frequency so that $A \Omega$ remains constant in the equation:
\begin{equation}
N(\lambda) =\frac{A\Omega}{\lambda^2} = A \Omega \left(\frac{\nu}{c}\right)^{2}.
\label{eq:multimode}
\end{equation} 
The detectors must provide broadband sensitivity over the full observing band. Their performance is parametrised by the noise equivalent power (NEP), $N_{\rm det}$, which includes contributions from photon noise, Johnson noise, thermal fluctuation noise, and $1/f$ noise. In the photon-noise-limited regime, the instrumental sensitivity scales approximately as $\sqrt{N_{\mathrm{det}}}$.\\
Finally, the detector time constant determines the maximum scanning speed of the FTS and therefore constrains the duration of each IFG. It also affects the effective integration time per IFG and consequently influences the overall observing strategy.

\begin{table*}[t]
\centering
\caption{Summary of the instrumental and mission parameters adopted in the \textit{reference configuration} and the parameter ranges explored throughout this work.}
\label{tab:instrument_parameters}

\begin{tabular}{llll}
\toprule
\textbf{Description} &
\textbf{Parameter} &
\textbf{Reference value} &
\textbf{Explored range} \\
\midrule
\midrule
\multicolumn{4}{l}{\textbf{Instrumental parameters}} \\

Minimum frequency &
$\nu_{\rm min}$ &
30 GHz &
$[1,\,180]$ GHz \\

Maximum frequency &
$\nu_{\rm max}$ &
2 THz &
$[1,\,6]$ THz \\

Spectral resolution &
$\Delta\nu$ &
15 GHz &
$[1,\,100]$ GHz \\

Number of detectors per FPA &
$N_{\rm det}$ &
1 &
1 \\

Detector efficiency &
$\eta$ &
0.5 &
$[0.2,\,0.8]$ \\

Number of modes at $\nu_{\rm min}$ &
$N$ &
8 &
$[1,\,100]$ \\

Primary mirror diameter &
$D_{\rm prim}$ &
30 cm &
$[15,\,60]$ cm \\

Optical properties &
$(\epsilon_k,\,T_k, \mathcal{T}_k)$ &
$(1\%,\,4.5\,\mathrm{K})$ &
$\epsilon_k\in[0.01,\,5]\%$,\quad
$T_k\in[3,\,70]$ K,\quad 
$\mathcal{T}_k\in[0.8,\,0.999]$ \\

\addlinespace[0.8ex]
\cmidrule(l){1-4}
\multicolumn{4}{l}{\textbf{Mission parameters}} \\

Effective observing time &
$t_{\rm obs}$ &
2.8 yr &
$[1~\mathrm{month},\,4~\mathrm{yr}]$ \\

Observation direction &
$\vec{p}$ &
see~\cite{Coulon2026FOSSIL_SKY} &
--- \\

\bottomrule
\end{tabular}

\end{table*}


\section{Instrument sensitivity estimation}
\label{sec:sensitivity_model}

We now present the methodology used to estimate the sensitivity of a Martin--Puplett FTS-based instrument. Throughout this work, we assume that the instrument operates in the photon-noise-limited regime. This assumption is justified for a FOSSIL-like instrument \citep{Aghanim2026FOSSIL_general_paper}, for which the photon noise is typically about two orders of magnitude larger than the other instrumental noise contributions. In particular, the Noise Equivalent Power (NEP) of state-of-the-art Kinetic Inductance Detectors (KIDs), $\mathrm{NEP}_{\rm det}\sim10^{-20}$~W\,Hz$^{-1/2}$ \citep{Catalano2020_KIDs, Baselmans2022kids}, is well below the FOSSIL requirement of $\sim3\times10^{-17}$~W\,Hz$^{-1/2}$, making the detector noise negligible compared to the photon noise.

We start by presenting how the signal propagates through the FTS, in Sect. \ref{subsec:signal_formation} and then derive, from the sky signal, the power received by the detectors (Sect. \ref{subsec:detector_power}). The associated photon noise is evaluated from this power and is converted into an equivalent  sensitivity.

\subsection{Signal propagation in a Martin--Puplett FTS}
\label{subsec:signal_formation}

An incident electromagnetic field written as $\vec{E} = A\hat{x} + B\hat{y}$, with $\hat{x}$ and $\hat{y}$ denoting orthogonal linear polarisations, propagates through the optical system of a symmetric Martin--Puplett interferometer and recombines at the output ports. The electric fields will depend on the optical path difference $\delta$ and after the final reflection on the last mirror pair of the FTS, the beams reach the detectors \citep{kogut2012pixie}.
\\
Considering an incident wave coming from the left-hand side of the telescope, the electric fields in the left arm ($\vec{E}_{L}$) and the right arm ($\vec{E}_{R}$) of the instrument at the detectors can then be written as:

\begin{align}
\vec{E}_{L} = -iB \sin\left(\frac{2\delta\omega}{c}\right) \hat{x} - iA \sin\left(\frac{2\delta\omega}{c}\right) \hat{y}, \\
\vec{E}_{R} = A \cos\left(\frac{2\delta\omega}{c}\right) \hat{x} - B \cos\left(\frac{2\delta\omega}{c}\right) \hat{y}. 
\end{align}

So far, the incident radiation entering the right-hand input port of the FTS has been neglected. Assuming that the instrument is perfectly symmetric, both in the FTS itself and in the telescopes, the derivation for the right-hand beam is identical to that of the left-hand beam. We therefore define two incident electric fields, $\vec{E}^{\rm sky}$ and $\vec{E}^{\rm AR}$, corresponding to the sky signal and the absolute reference, entering the left and right input ports of the FTS, respectively.

By applying the principle of superposition, which allows contributions from both FTS inputs to be combined for all frequencies, the incident power received by each detector, for each polarisation state ($\hat{x}$ and $\hat{y}$) and for each output arm ($L$ and $R$), can be written as a function of the optical path difference $\delta$:

\begin{align}
    P_{Lx} &= \frac{1}{2}  \int \left[ \left( E^{{\rm sky}^2}_y + E^{{\rm AR}^2}_x \right) +  \left( E^{{\rm AR}^2}_x - E^{{\rm sky}^2}_y \right) \cos \left( \frac{4 \delta  \omega}{c}\right) \right] d\omega, \\
    P_{Ly} &= \frac{1}{2}  \int \left[ \left( E^{{\rm sky}^2}_x + E^{{\rm AR}^2}_y \right) +  \left( E^{{\rm AR}^2}_y - E^{{\rm sky}^2}_x \right) \cos \left( \frac{4 \delta  \omega}{c}\right) \right] d\omega,\\
    P_{Rx} &= \frac{1}{2}  \int \left[ \left( E^{{\rm sky}^2}_x + E^{{\rm AR}^2}_y \right) +  \left( E^{{\rm sky}^2}_x - E^{{\rm AR}^2}_y \right) \cos \left( \frac{4 \delta  \omega}{c}\right) \right] d\omega, \\
    P_{Ry} &= \frac{1}{2}  \int \left[ \left( E^{{\rm sky}^2}_y + E^{{\rm AR}^2}_x \right) +  \left( E^{{\rm sky}^2}_y - E^{{\rm AR}^2}_x \right) \cos \left( \frac{4 \delta  \omega}{c}\right) \right] d\omega
\end{align}
The previous expressions give the incident power on each detector for a given output arm and polarisation state. For detectors insensitive to polarisation, the measured signal is obtained by summing the $\hat{x}$ and $\hat{y}$ contributions. Rewriting these expressions in terms of the Stokes parameters, the power on detectors become:
\begin{align}
P_{L} &= \frac{1}{2}\int  I_{\mathrm{\rm sky}} + I_{\mathrm{\rm AR}} + \left( I_{\mathrm{\rm sky}} - I_{\mathrm{\rm AR}} \right)
\cos\left(\frac{4\delta\omega}{c}\right)\mathrm{d}\omega, \\
P_{R} &= \frac{1}{2}\int  I_{\mathrm{\rm sky}} + I_{\mathrm{\rm AR}} +  \left( I_{\mathrm{\rm AR}} - I_{\mathrm{\rm sky}} \right)
\cos\left(\frac{4\delta\omega}{c}\right)\mathrm{d}\omega.
\end{align}

\begin{figure}[!ht]
    \centering
    \includegraphics[width=7cm]{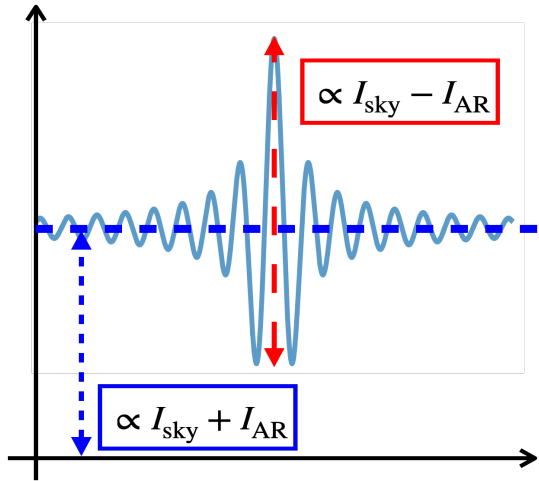}
    \caption{Example of an interferogram obtained with a FOSSIL-like instrument. Its average level (blue) is proportional to the frequency-integrated sum of the sky and absolute reference intensities, whereas the modulated component (red) is proportional to their frequency-integrated difference.}
    \label{fig: IFG}
\end{figure}

In the FOSSIL configuration, the detector bandwidth are divided into two spectral sub-bands using a dichroic element placed at the FTS outputs.  In each sub-band, the detectors record an IFG as the scanning mirror mechanism varies the optical path difference, $\delta$, between the two interferometer arms. The measured power consists of an unmodulated component and an interferometric component proportional to the difference between the sky and absolute reference signals as shown in Fig.\ref{fig: IFG}.

This measurement simultaneously provides an absolute measurement, through the presence of the absolute reference, and a differential measurement between the sky signal and the reference signal. It is therefore sensitive to deviations from the spectrum of the absolute reference. Consequently, if the latter corresponds to a blackbody at the CMB temperature, the instrument can directly probe CMB spectral distortions. In practice, this requires the subtraction of all contaminating contributions from the detector signal, including astrophysical emissions and the instrument's own internal emission.

The full derivation of the electromagnetic wave propagation through the Martin--Puplett FTS, together with the calculation of the power reaching the detectors are detail in \cite{LeGall2026FOSSIL_optics}.
 
\subsection{Optical power at the detectors}
\label{subsec:detector_power}

In the following, we derive the expression of the optical power received by the detectors in an idealised instrument configuration.
\\
The signal measured by the  detectors is dominated by sky emission, but also includes thermal emission from the instrument itself. We assume that only optical elements contribute significantly to the thermal emission, while internal reflections are neglected. As seen in Sect. \ref{subsubsec:optics}, each optical component is modelled as a grey body and the emitted spectral radiance is given by
\begin{equation}
B(\nu,T_k,\epsilon_k) = \epsilon_k(\nu)\, B(\nu,T_k)
= \frac{2 h \nu^3}{c^2}\,\frac{\epsilon_k(\nu)}{\exp\!\left(\frac{h\nu}{k_B T_k}\right)-1}.
\end{equation}

The radiation emitted by a given element propagates towards the detectors and is transmitted by all subsequent optical elements. Neglecting reflections, the total power received by the detectors can be written as:
\begin{equation}
P =
S_{\mathrm{\rm tot}}\, \mathcal{T}_{\rm tot}(\nu)
+
\sum_{k=1}^{K} \epsilon_k(\nu)\, B(\nu,T_k)\,
\prod_{j=1}^{k-1} \mathcal{T}_j(\nu),
\label{eq:total_power_sky_instrument}
\end{equation}
where $\mathcal{T}_{\rm tot}(\nu)=\prod_{k=1}^{K} \mathcal{T}_k(\nu)$ denotes the total optical transmission of the instrument, and $S_{\mathrm{tot}} = S_{\mathrm{sky}} + S_{\mathrm{AR}}$ is the total incident signal entering the instrument, composed of the sky emission and the emission from the absolute reference blackbody.

The corresponding spectral power at each detector output is thus
\begin{equation}
P_{\nu}(\nu)=
\frac{1}{2}\,\eta(\nu)\,
\left[
S_{\mathrm{\rm tot}} \prod_{k=1}^{K} \mathcal{T}_k(\nu)
+
\sum_{k=1}^{K} \epsilon_k(\nu)\, B(\nu,T_k)\prod_{l=1}^{k-1} \mathcal{T}_l(\nu)
\right]
A\,\Omega(\nu).
\label{eq:spectral_power_detector}
\end{equation}

This expression corresponds to a conservative case in which the optical path difference is zero, so that all frequency components are added incoherently at the detector level. This assumption provides an upper bound on the photon noise and therefore a conservative estimate of the instrument sensitivity.

\subsection{Noise Equivalent Power}
\label{subsec:NEP}

The sensitivity of a bolometric detector is ultimately limited by its Noise Equivalent Power (NEP), which quantifies the minimum detectable optical power in the presence of noise. The total NEP includes contributions from photon noise, detector noise, readout noise, and other instrumental effects.

For a FOSSIL-like instrument, we assume a photon-noise-limited regime, in which the photon NEP can be written as:
\begin{equation}
\mathrm{NEP}_{\mathrm{photon}}^2 =
2 \int_{\nu_{\mathrm{min}}}^{\nu_{\mathrm{max}}}
\left(
h\nu\, P_{\nu}
+
\frac{c^2}{\nu^2 A\Omega}\, P_{\nu}^2
\right)\,\mathrm{d}\nu,
\label{eq:nep_photon}
\end{equation}
where $P_{\nu}$ is given by Eq.~\ref{eq:spectral_power_detector}. Similar formulation can be found for example in \cite{Mather1982_BolometerNoise} and \cite{Lamarre1989}.

\subsection{Sky sensitivity}
\label{subsec:sensitivity_sky}

The NEP at detector level can be converted into an equivalent sky brightness sensitivity. Following the approach of \cite{kogut2012pixie}, the total noise on the measured power after an integration time $t_{\rm obs}$ writes
\begin{equation}
\delta P = \frac{\sqrt{2}\,\mathrm{NEP}_{\mathrm{tot}}}{\sqrt{ t_{\rm obs}}},
\label{eq:deltaP}
\end{equation}
where the factor of $\sqrt{2}$ accounts for the conversion between time and frequency domains in the Fourier-transform measurement and the splitting of the signal in the two arms of the instrument. 

The effective integration time depends on the sampling rate of both the scanning mirror position and the detectors. Assuming that the FTS sampling is matched to the detector time constant, the effective integration time becomes equal to the observing time. Under this assumption, the effective integration time can be derived directly from the nominal duration of the mission. For a FOSSIL-like mission, we assume that 70\% of the nominal mission duration is effectively dedicated to sky observations, corresponding to a effective observing time of 2.8 years.

The noise at the detector can then be converted into a specific intensity in the sky:
\begin{equation}
\delta I_{\nu} =
\frac{\delta P}{A\,\Omega\, \Delta \nu \, \eta\, \mathcal{T}_{\rm tot} \sqrt{N_{\mathrm det}}},
\label{eq:sensitivity_sky}
\end{equation}
where $\Delta \nu$ is the effective spectral bandwidth.\\
This quantity represents the minimum detectable sky intensity per frequency channel. A sky signal equal to $\delta I_{\nu}$ would therefore correspond to a signal-to-noise ratio of one.

Several key relations follow directly from this expression. Increasing the optical throughput $A\Omega$ raises both the signal and the photon noise; however, the signal increases linearly while the noise increases only as $(A\Omega)^{1/2}$, leading to a net improvement in sensitivity proportional to $(A\Omega)^{1/2}$. Similarly, the sensitivity improves as $\sqrt{t_{\rm obs}}$ and $\sqrt{N_{\mathrm{det}}}$, justifying the use of detector arrays and long integration times. 
In the following, when referring to the instrument sensitivity, we will implicitly consider this noise limit $\delta I_{\nu}$, i.e.  the sky intensity corresponding to a signal-to-noise ratio of one.


\section{Instrument optimisation}
\label{sec:app_case}

The optimisation of an instrument for the detection or measurement of a given astrophysical signal consists in identifying, for a predefined instrument concept, the set of parameters that maximises the scientific return while satisfying instrumental and mission constraints. This is particularly challenging for measurements targeting extremely faint signals whose detectability depends not only on the instrument characteristics but also on foreground contamination and systematic effects.

Here, we consider the optimisation of an FTS concept designed to measure CMB spectral distortions, such as the one proposed for the FOSSIL mission. More specifically, we investigate how the instrumental parameters affect the measurement of the $y$- and $\mu$-type distortions monopole, together with the CIB amplitude $A_{\rm CIB}$, namely, assessing what is the impact on the forecasted SNR.

The optimisation is intrinsically a high-dimensional problem, and the scientific performance, expressed as the signal-to-noise ratio (SNR) on the astrophysical parameters, results from the interplay between instrumental parameters, mission characteristics, and sky-model assumptions. These quantities are often strongly correlated and, in some cases, degenerate, preventing a purely intuitive or fully analytical assessment of their combined impact. In addition, while the influence of individual parameters on the instrumental sensitivity could be understood qualitatively, their propagation to the scientific performance is considerably more complex. Therefore we present two complementary approaches that were developed to address this issue. The first consists of a one-parameter-at-a-time optimisation strategy (Sect. \ref{sec:one-param}), in which each instrumental parameter is varied independently while all others are kept fixed. The second is a global optimisation method based on ML (Sect. \ref{sec:ML-param}).

\subsection{One-parameter-at-a-time optimisation}\label{sec:one-param}
\begin{figure*}[!t]
    \centering
    \includegraphics[width=\linewidth]{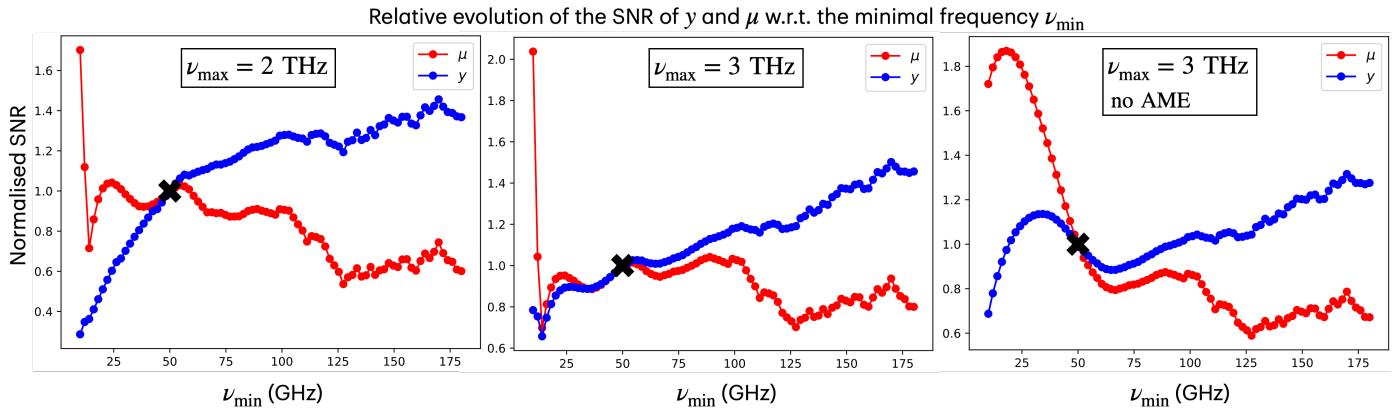}
    \caption{Evolution of the SNR of the $y$-type (blue) and $\mu$-type (red) spectral distortion monopoles as a function of the minimum observing frequency, $\nu_{\mathrm{min}}$, for a given maximum frequency $\nu_{\mathrm{max}}$ of $2$~THz (left) and 3~THz (middle). The right panel is as as the middle one with a sky model in which the AME emission is neglected.
    The SNR value (black cross) used for normalisation correspond to the reference instrument configuration  Table~\ref{tab:instrument_parameters}.
    }
    \label{fig:f_min_trend}
\end{figure*}

To illustrate this optimisation approach, we consider the case of minimum observing frequency, $\nu_{\mathrm{min}}$, of a FOSSIL-like instrument, which is varied while all other parameters are kept fixed. By construction, this approach largely neglects parameter correlations and explores only a limited fraction of the full multidimensional parameter space. The forecasted scientific performance, i.e. SNRs of the astrophysical parameters ($y$ and $\mu$), are obtained using a Fisher matrix formalism under the assumption of Gaussian posteriors. For this purpose, the considered sky model is the one summarised in ~Sect.\ref{sec:1-sky} and the reference instrumental configuration is the one described in Sect.~\ref{sec:instrument_concept}. The complete forecasting pipeline that combines the sky model with the instrument model is detailed in \cite{Coulon2026FOSSIL_SKY}. 
\\
In practice, $\nu_{\mathrm{min}}$ is varied between $10$~GHz and $180$~GHz with a step of $2$~GHz, and the SNRs for both the $y$ and $\mu$ distortions are computed with respect to the reference FOSSIL configuration summarised in Table~\ref{tab:instrument_parameters}. The resulting relative SNR variations are shown in Fig.~\ref{fig:f_min_trend} for three different cases. In the left panel, the maximum observing frequency is fixed to $\nu_{\mathrm{max}}=2$~THz which  corresponds to the reference configuration of FOSSIL, while in the middle and right panels it is fixed to $\nu_{\mathrm{max}}=3$~THz. The right panel additionally considers a sky model in which the AME emission is neglected. In all three cases, the black cross represents the reference configuration minimal frequency, while the blue and red lines show the relative SNR for the $y$ and $\mu$ distortions, respectively.

Decreasing the minimum observing frequency, $\nu_{\rm min}$, significantly improves the SNR of the $\mu$-distortion monopole, enabling an enhancement of up to $70\%$ with respect to the reference configuration at $\nu_{\rm min}=10$~GHz. The improvement factor is not monotonic and exhibits several local maxima and minima. These features arise from the interplay between the characterization of AME and the instrument characteristics. Since the spectral resolution is fixed to 15~GHz, changing $\nu_{\rm min}$ shifts the frequencies sampled by the first few channels, thereby modifying their ability to constrain the AME spectrum. As low-frequency channels become available, the AME constraint improves thereby increasing the $\mu$ SNR, whereas extending the frequency coverage also reduces the sensitivity per channel (Eq.~\ref{eq:sensitivity_sky}), leading to decreases in SNR.

In contrast, the $y$-distortion monopole exhibits an opposite behaviour: its SNR increases nearly monotonically with increasing $\nu_{\rm min}$, reaching an improvement of about $40\%$ at $\nu_{\rm min}=170$~GHz. 

The dependence on the instrumental configuration is also illustrated by increasing the maximum observing frequency from $\nu_{\rm max}=2$~THz to $3$~THz while keeping all other parameters unchanged (middle panel of Fig.~\ref{fig:f_min_trend}). The same overall trends are recovered: the $\mu$ SNR still favours lower values of $\nu_{\rm min}$, whereas the $y$ SNR favours higher ones. The main differences are a reduced overall variation for the $y$ distortion (from about $110\%$ to $90\%$ relative to the reference configuration) and a much stronger enhancement of the $\mu$ SNR below $\nu_{\rm min}\simeq14$~GHz, where it reaches nearly $200\%$ . In addition, the oscillatory behaviour associated with the varying constraining power on AME becomes visible for both spectral distortion monopoles.

This interpretation is further supported by repeating the analysis after removing the AME contribution from the sky model \citep[see][]{Coulon2026FOSSIL_SKY}. Above $\nu_{\rm min}\simeq80$~GHz, the behaviour is essentially unchanged. At lower frequencies, however, the SNR curves become significantly smoother and the pronounced enhancement of the $\mu$ SNR below $\sim14$~GHz disappears. This demonstrates that both the oscillatory behaviour and the strong low-frequency enhancement are primarily driven by the ability of channels near the AME emission peak to constrain this foreground and partially lift its degeneracy with the spectral distortions. Small residual oscillations nevertheless remain, indicating that a similar, although weaker, mechanism is still at work and associated with the characterization of the other low-frequency foregrounds, namely synchrotron and free-free emissions.

In the absence of AME, the SNR evolution is therefore mainly governed by the competition between the increased spectral information provided by extending the frequency coverage towards lower frequencies and the associated loss of instrumental sensitivity. Both SNR curves first decrease to a local minimum around $\nu_{\rm min}\simeq64$~GHz, then increase to maxima at $\nu_{\rm min}\simeq18$~GHz for $\mu$ and $\nu_{\rm min}\simeq32$~GHz for $y$, corresponding to improvements of more than $85\%$ and about $15\%$, respectively, before decreasing again towards lower frequencies.

\subsection{ML-based optimisation}\label{sec:ML-param}

The previous optimisation strategy, based on varying one instrumental parameter at a time, all others being fixed, provides valuable insight into the individual impact of the different instrument characteristics without capturing all the possible correlations between instrument parameters. Therefore, it is not efficient to identify the trades-off that lead to a global optimum instrumental configuration. Rather than resorting to a brute-force exploration of the complete parameter space requiring the evaluation of a very large number of instrumental configurations, we develop a global optimisation framework based on ML. 
\\
In our context, the regression model must be capable of capturing non-linear relationships, remain stable under perturbations, generalise well outside the training set, and provide a sufficient level of interpretability. Computational efficiency of the ML model is also important given the large number of evaluations required. The above mentioned requirements converge towards the use of the particularly well suited decision tree-based methods, as they naturally handle non-linear dependencies and parameter interactions while remaining relatively efficient to train and interpret.

In the present study, we use and compare the two ML models described hereafter. In both cases, the procedure starts by defining the ranges of variation of all the instrumental parameters listed in Table~\ref{tab:instrument_parameters}, except the number of detectors per focal plane, $N_{\rm det}$, and the optical transmission, $\mathcal{T}_k$. These ranges are then sampled to generate a dataset of instrumental configurations, denoted by $\Theta_{\rm instru}=\{\theta_{\rm instru}\}$, where each configuration $\theta_{\rm instru}$ is a vector containing the values of all instrumental and mission parameters. 

Based on the allowed ranges of variation defined for each parameter of the instrument model (see Table \ref{tab:instrument_parameters}), we generate $\Theta_{\rm instru}$, a sample of $100,000$ instrumental configurations. To limit the computational cost of the analysis while ensuring an efficient quasi-random sampling of the parameter space, the configurations are generated using Latin hypercube sampling \citep{McKay1979Latin_HC_sampling,Stein1987Latin_HC_sampling}. This method provides a quasi-random distribution of samples over a multidimensional parameter space $\Omega$, while ensuring that each sample occupies a unique interval along each dimension, thereby avoiding redundant sampling of nearly identical configurations.

Next, for each configuration $\theta_{\rm instru}$, the forecasting procedure described in \cite{Coulon2026FOSSIL_SKY} is applied to compute the corresponding vector of forecasted SNRs for the sky-model parameters, $\mathcal{F}(\theta_{\rm instru})=\sigma_{\rm sky}$, where $\mathcal{F}$ denotes the forecasting function and $\sigma_{\rm sky}$ the corresponding vector of predicted SNRs. Unlike in \cite{Coulon2026FOSSIL_SKY}, no external priors are applied to the sky-model parameters in the present analysis. Repeating this procedure for every configuration in $\Theta_{\rm instru}$ produces the complete forecast dataset, $\mathcal{F}(\Theta_{\rm instru})=\Sigma_{\rm sky}$, where $\Sigma_{\rm sky}$ denotes the set of all forecasted SNR vectors. 

A decision-tree regression model is then trained on 85\% of the paired dataset $(\Theta_{\rm instru}^{\rm train},\Sigma_{\rm sky}^{\rm train})$, while the remaining 15\% is reserved for validation. The objective is to learn the forecasting function
$\mathcal{F}:\theta_{\rm instru}\rightarrow\sigma_{\rm sky}$, thereby providing fast predictions of the forecasted SNRs for any instrumental configuration.

\subsubsection{Random Forest: formalism}
\label{subsubsec:rf}

Random Forest (RF) models are based on an ensemble of decision trees trained on bootstrap samples of the data and random subsets of the input features. The final prediction is obtained by averaging the predictions of all trees, which reduces the variance of individual trees while preserving their ability to capture highly non-linear relationships \citep{breiman2001random,hastie2009random,scornet2015consistency}. Random Forest regressors have already proven effective for astrophysical applications \citep{Richards2012RF,CarrascoKind2013RF,Kim2015RF, Dai2018RF, FlukeJacobs2020RF,ArroquiaCuadros2023RF}. Besides providing accurate regression models with little data preprocessing, RFs naturally quantify the relative importance of each input feature through the reduction in prediction error induced by successive tree splits, making them particularly well suited for global sensitivity analysis \citep{biau2012analysis}.

In this work, we use the \texttt{RandomForestRegressor} implementation available in \texttt{scikit-learn} \citep{sklearn_pedregosa2011}. For each predicted sky-parameter SNR ($y$, $\mu$, and $A_{\rm CIB}$), an independent RF model, denoted $F_{\rm RF}$, is trained using the same training and validation datasets described above. The forests are composed of 50 decision trees for each regression, while all remaining hyperparameters are left to their default \texttt{scikit-learn} values. A fixed random seed (\texttt{random\_state}=42) is adopted to ensure reproducibility. Although the RF acts as an emulator of the forecasting function $\mathcal{F}$, its primary purpose in this work is to perform a global sensitivity analysis. To this end, the relative importance of each instrumental parameter is derived from the impurity-based feature importance provided by the trained forest, while its uncertainty is estimated from the standard deviation of the feature importance across the individual trees.

After training the RF regressor on the paired training dataset $(\Theta_{\rm instru}^{\rm train},\Sigma_{\rm sky}^{\rm train})$, the relative importance, $I_i$, of each instrumental parameter is estimated using the impurity-based feature importance implemented in \texttt{scikit-learn}. We denote the vector of $p$ instrumental parameters by $\theta_{\rm instru}=\{\theta_1,\ldots,\theta_p\}$, where $\theta_i$ represents the $i$-th instrumental parameter. The importance of a given parameter $\theta_i$ is defined as the normalized total reduction in regression impurity (variance) obtained from every node split across all trees in the forest that uses $\theta_i$ as the splitting parameter,
\begin{equation}
I_i=\frac{\Delta_i}{\sum_{j=1}^{p}\Delta_j},
\end{equation}
where $\Delta_i$ is the cumulative impurity decrease attributed to parameter $\theta_i$. By construction,
\begin{equation}
\sum_{i=1}^{p} I_i = 1.
\end{equation}

The predictive performance of the RF emulator is evaluated on the validation dataset $(\Theta_{\rm instru}^{\rm val},\Sigma_{\rm sky}^{\rm val})$ through the coefficient of determination, defined as
\begin{equation}
R^2 =
1-
\frac{
\displaystyle\sum_{\theta_{\rm instru}\in\Theta_{\rm instru}^{\rm val}}
\left[
\mathcal{F}(\theta_{\rm instru})
-
F_{\rm RF}(\theta_{\rm instru})
\right]^2
}{
\displaystyle\sum_{\theta_{\rm instru}\in\Theta_{\rm instru}^{\rm val}}
\left[
\mathcal{F}(\theta_{\rm instru})
-
\overline{\mathcal{F}}
\right]^2
},
\label{eq: R2}
\end{equation}
with the mean forecasted SNR over the $N_{\rm val}$ instrumental configurations of the validation dataset, $\Theta_{\rm instru}^{\rm val}$ defined as:
\begin{equation}
\overline{\mathcal{F}}
=
\frac{1}{N_{\rm val}}
\sum_{\theta_{\rm instru}\in\Theta_{\rm instru}^{\rm val}}
\mathcal{F}(\theta_{\rm instru}),
\end{equation}
where, $\mathcal{F}(\theta_{\rm instru})$ denotes the forecasted SNR obtained from the full forecasting pipeline for the instrumental configuration $\theta_{\rm instru}$, while $F_{\rm RF}(\theta_{\rm instru})$ is the corresponding prediction of the RF emulator. \\
The coefficient of determination, $R^2$, measures the fraction of the variance of the forecasted SNRs explained by the RF emulator. A coefficient of one means that the RF predictions match perfectly the actual data points.

%
\subsubsection{Random Forest: results}
\label{subsubsec:rf_results}

\begin{figure}[!ht]
    \centering
    \includegraphics[width=9cm]{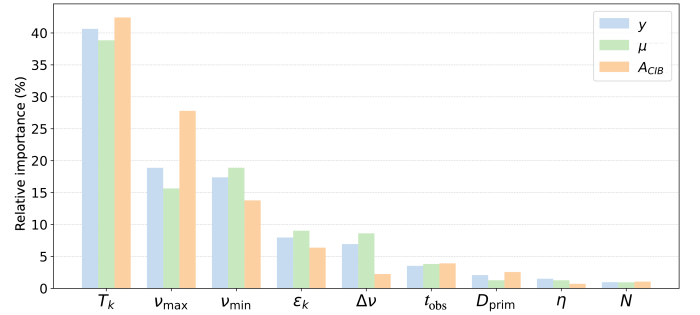}
    \caption{Relative importance of instrumental parameters for the forecasted measurements of the $y$-distortion monopole (blue), the $\mu$-distortion monopole (green), and the CIB monopole amplitude (orange).}
    \label{fig: RF}
\end{figure}

Before analysing the outputs of the trained RF model, we estimate its performance in terms of the coefficients of determination for our target sky parameters, namely $y$, $\mu$, and $A_{\mathrm{CIB}}$. We find $0.870$, $0.777$, and $0.879$, respectively, indicating that the RF emulator provides a good approximation of the forecasting function. The lowest performance is obtained for $\mu$. This is mainly because, for a significant fraction of the sampled instrumental configurations, the $\mu$-distortion signal is smaller than the instrumental sensitivity, corresponding to forecasted SNRs of order unity. In this regime, the forecasted SNR spans a relatively narrow range and varies only weakly with the instrumental parameters, making the regression task more challenging for the RF model.
Figure~\ref{fig: RF} presents the relative importance of the instrumental parameters for the three observables.

Despite observable-dependent variations, a consistent set of dominant parameters emerges across all cases: $\nu_{\min}$, $\nu_{\max}$, $T_k$, $\Delta \nu$ and $\epsilon_k$. These parameters therefore primarily drive the sensitivity of the instrument to CMB spectral distortion and foreground measurements. The maximum frequency $\nu_{\max}$ has a stronger impact on $A_{\mathrm{CIB}}$ than on CMB spectral distortions, while $T_k$ affects all observables at a comparable level. Given the obtained results, it seems that the spectral resolution $\Delta \nu$ plays a more important role for CMB spetral distortions and in particular for $\mu$, perhaps reflecting its lower intrinsic amplitude. \\
By contrast, $N$, $D_{\mathrm{prim}}$, $\eta$ and $t_{\mathrm{obs}}$ exhibit low subdominant impact, each contributing less than $5\%$ to the total variance. Combined together, they account for at most $\sim10\%$ of the overall sensitivity variations. The comparable influence of $N$, $D_{\mathrm{prim}}$, and $\eta$ is consistent with the scaling of the instrumental sensitivity (see Eq.~\ref{eq:sensitivity_sky}).

With Random Forest models, we find that the relative importance of the input parameters remains only moderately constrained. For the $\mu$ distortion, the model achieves a score of 0.777, indicating that a significant fraction ($\geq 22\%$) of the variance is not captured. This is reflected in the dispersion of feature importance across the ensemble of decision trees that reaches $\sim20\%$ for the most influential parameters and up to $100\%$ for the least significant ones.

\subsubsection{Gradient Boosting: formalism}
\label{subsubsec:gb}

In contrast to RF methods, where the individual decision trees are trained independently on different bootstrap samples of the data, Gradient Boosting \citep{Friedman2001_CML_boosting} combines decision trees sequentially to construct a predictive model. At each iteration, a new tree is fitted to the errors of the current model, more precisely to the negative gradient of the chosen loss function with respect to the current predictions. For a squared-error loss, this corresponds to fitting the new tree to the residuals, i.e. to the differences between the target values and the predictions of the current ensemble \citep{Friedman2002_stochastic_boosting}. The prediction is then progressively updated by adding the contribution of each new tree, with the overall model being optimised by gradient descent to minimise the chosen loss function. This differs fundamentally from Random Forests, for which the final prediction in regression is obtained by averaging the predictions of independently trained trees. Although this averaging provides an effective and robust ensemble prediction, the resulting RF model can be affected by noise or biases in variable-importance estimates \citep{Strobl2007_BiasRFVarImp,Strobl2008_CondVarImpRF}. By contrast, the sequential optimisation of Gradient Boosting allows the model to progressively adapt to the structure of the data and to capture complex non-linear dependencies and interactions between the instrumental parameters \citep{Friedman2001_CML_boosting}. This results in a more accurate approximation of the underlying response and, consequently, in more stable estimates of the relative importance of the input parameters. The increased flexibility of the model, however, also makes Gradient Boosting more susceptible to overfitting and makes the choice of its hyperparameters, such as the number of estimators, learning rate, and tree complexity, particularly important \citep{Caruana2006_SupervisedLearningComparison}.

In this work, we use the \texttt{XGBoost} implementation of gradient boosting \citep{xgboost}. To interpret the resulting model, we further use \textit{SHapley Additive exPlanations} (SHAP) values \citep{shap}. 
SHAP provides a general framework for attributing the output of a ML model to its input features, each SHAP value quantifies the contribution of an individual feature to the difference between the model prediction and a reference value. These contributions are based on Shapley values from cooperative game theory \citep{Shapley1953_ShapleyValue, Winter2002_ShapleyValue} and are obtained by evaluating the marginal contribution of a feature across different subsets of input features \citep{Marzouk2025_SHAPtractability}. This provides a consistent decomposition of the model prediction into individual feature contributions while accounting for the different combinations in which features can contribute to the prediction. For tree-based models such as \texttt{XGBoost}, this attribution can be computed using the \textit{TreeSHAP} algorithm \citep{Lundberg2018_TreeSHAP}.

\subsubsection{Gradient Boosting: results}
\label{subsubsec:gb_results}

\begin{figure*}[!ht]
    \centering
    \includegraphics[height=6.3cm]{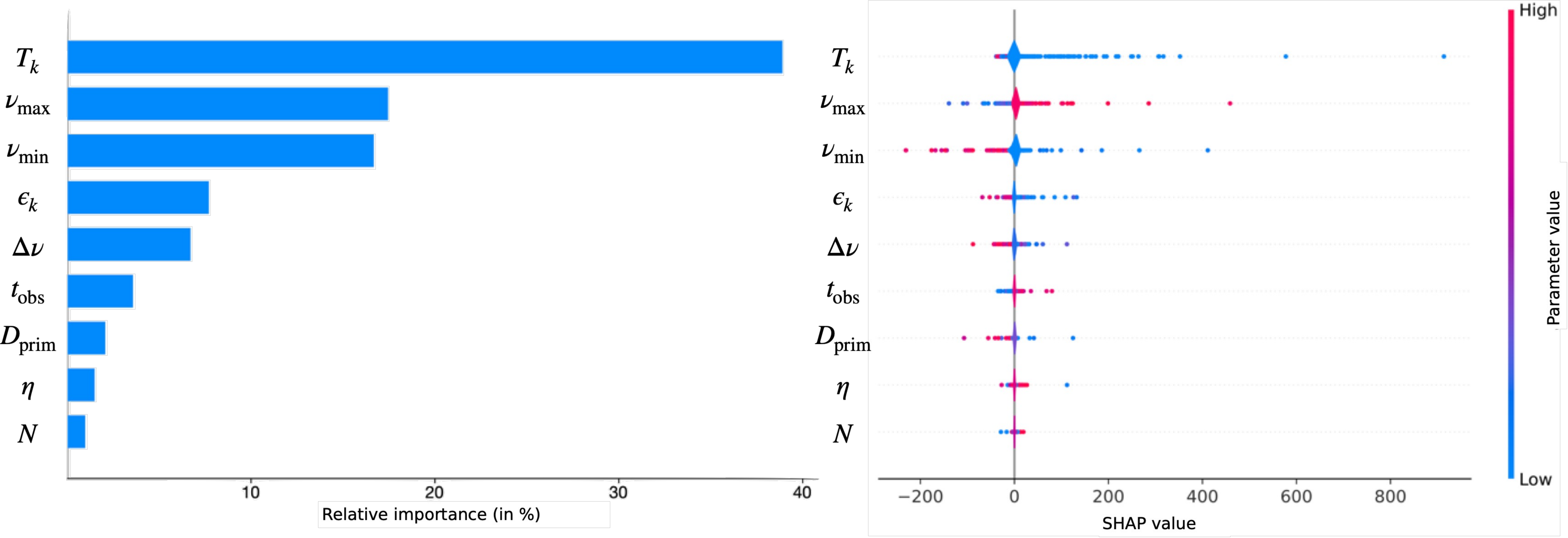}
    \caption{\textit{Left}: Mean SHAP-based relative importance of instrumental parameters for the prediction of the $y$-SNR. \textit{Right}: SHAP values for individual realisations. The horizontal displacement from zero quantifies the impact of each parameter on the predicted $y$-SNR. Positive (negative) values indicate an increase (decrease) relative to the mean prediction corresponding to a SHAP value of 0. Colours encode the parameter value within its prior range, from low (blue) to high (red).}
    \label{fig: SHAP y}
\end{figure*}

In the following, SHAP values are used to characterise how the instrumental parameters affect the predicted measurement capabilities on the three observables considered in this study, namely $y$, $\mu$, and $A_{\rm CIB}$. The relative importance of each parameter can be estimated from the mean absolute SHAP value over the sample, providing a global measure of its contribution to the model predictions. In addition, the individual SHAP values retain information on the direction of the effect. A positive SHAP value indicates that a given parameter contributes to increasing the predicted SNR relative to the mean model prediction, whereas a negative value indicates a contribution towards decreasing it. The magnitude of the SHAP value quantifies the strength of this contribution. This analysis therefore allows us to identify not only which instrumental parameters have the strongest influence on the predicted sensitivity, but also how their values affect the model prediction, as illustrated in Fig.~\ref{fig: SHAP y}.

For all three observables, the \texttt{XGBoost} model achieves a higher coefficient of determination (see Eq.~\ref{eq: R2}) than the RF model, as shown in Table~\ref{tab:R2_RF_XGBoost}. The improvement is observed consistently for $y$, $\mu$, and $A_{\mathrm{CIB}}$, with the $R^2$ increasing from $0.870$ to $0.927$, from $0.777$ to $0.891$, and from $0.879$ to $0.933$, respectively. On average, \texttt{XGBoost} method increases the coefficient of determination by 7.5\% relative to the RF model. In other words, $F_{\mathrm{XGB}}$ recovers, on average, $7.5\%$ more of the variance of $\mathcal{F}$ than $F_{\mathrm{RF}}$. The largest improvement is obtained for $\mu$, with an increase of 11.4\% in $R^2$, while the improvements for $y$ and $A_{\mathrm{CIB}}$ are 5.7\% and 5.4\%, respectively.

\begin{table}[htbp]
    \centering
    \caption{Comparison of the coefficient of determination, $R^2$, obtained with the RF and \texttt{XGBoost} models for the parameters $y$, $\mu$, and $A_{\mathrm{CIB}}$.}
    \label{tab:R2_RF_XGBoost}
    \begin{tabular}{lcc}
        \hline
        Target parameter & Random Forest & \texttt{XGBoost} \\
        \hline
        $y$             & $0.870$ & $0.927$ \\
        $\mu$           & $0.777$ & $0.891$ \\
        $A_{\mathrm{CIB}}$ & $0.879$ & $0.933$ \\
        \hline
    \end{tabular}
\end{table}
The \texttt{XGBoost} model also enables a more detailed analysis of how the instrumental parameters affect the measurement of the three observables using SHAP values, as illustrated in the SHAP analysis for $y$ (Fig.~\ref{fig: SHAP y}).

The temperature of the warmest optical component, $T_k$, is identified as the dominant parameter governing the sensitivity to $y$-type distortions. The frequencies $\nu_{\min}$ and $\nu_{\max}$ exhibit impacts approximately a factor of two lower, followed by $\epsilon_k$ and the spectral resolution.
The SHAP distribution shows a clear and systematic dependence on $T_k$. High temperatures systematically degrade performance, while low temperatures lead to strongly positive contributions. Quantitatively, large positive SHAP values occur only for $T_k \lesssim 10\,\mathrm{K}$, with extreme contributions for $T_k \lesssim 5\,\mathrm{K}$. This confirms the significant gain in sensitivity achieved when operating the warmest component at temperatures of order $4.5\,\mathrm{K}$, as done for the FOSSIL mission \citep{Aghanim2026FOSSIL_general_paper}. A similar behaviour is observed for $\nu_{\min}$ and $\nu_{\max}$, confirming that a wide spectral range improves sensitivity to $y$-type distortions.

The SHAP analysis for $\mu$ (see Fig.\ref{fig: SHAP mu}) shows that the minimum frequency $\nu_{\min}$ plays a more significant role than $\nu_{\max}$, consistent with the lower characteristic frequency of the $\mu$-distortion spectrum. The emissivity, $\epsilon_k$, and spectral resolution, $\Delta \nu$, also contribute more strongly than in the $y$ case, reflecting the reduced signal amplitude. The overall ranking of the remaining parameters, and the impact of their variations, remain similar.
For $A_{\mathrm{CIB}}$ (see Fig.\ref{fig: SHAP cib}), the dominant parameters are $T_k$, $\nu_{\max}$, and $\nu_{\min}$. However, the impact of $\nu_{\max}$ is approximately twice that of $\nu_{\min}$, in contrast with the near-equal contributions observed for $y$ and $\mu$. This is consistent with the high-frequency nature of the CIB signal. The spectral resolution has a comparatively weak impact on $A_{\mathrm{CIB}}$, as coarse sampling is sufficient to resolve the spectral shape in the high-frequency regime. 
The corresponding SHAP figures for the $\mu$ and $A_{\mathrm{CIB}}$ parameters are presented in Appendix~\ref{app:shap}.


\section{Discussion}
\label{sec: discu}

The following discussion addresses the robustness and validity of the ML approach, the trade-off between sensitivity and frequency coverage, and some limitations of the adopted instrument and sky models.


\subsection{Robustness and validity of the ML methodology}
\label{subsec: robustness_global}

The consistency of the results obtained with the two machine-learning approaches provides an indication of the robustness of the identified instrumental dependencies. Both the RF and \texttt{XGBoost} models recover similar dominant parameters while the latter provides a higher coefficient of determination for all three observables. This indicates that the main trends identified in the parameter space are not specific to a particular decision tree-based algorithm. The higher predictive accuracy of \texttt{XGBoost} then allows a more detailed interpretation of these dependencies through the SHAP analysis, which reveals a small number of apparently non-physical feature contributions.

In particular, for the detector efficiency, $\eta$, an isolated point associated with low efficiency exhibits a large positive SHAP contribution to the predicted $y$-SNR (Fig.~\ref{fig: SHAP y}). This behaviour is not physically expected, as decreasing the detector efficiency cannot improve the instrumental sensitivity. This raises the question of whether the non-physical behaviour originates from the \texttt{XGBoost} model or from the SHAP analysis itself. We performed several tests and diagnostics to assess the predictive accuracy and robustness of the  \texttt{XGBoost} model (see Appendix~\ref{app:robustness}). The isolated non-physical SHAP contribution can therefore be interpreted as a local limitation of the learned model rather than a systematic failure of the method. If the model locally approximates the physical relation inaccurately, the resulting SHAP values can also become locally non-physical, even when the model has good predictive performance overall.

As is the case in other ML usage, we find that the robustness of our conclusions is conditional on the underlying physical model. The ML models are trained on a finite set of configurations generated using a simplified description of the instrument and sky emission. They can therefore reproduce only the physical dependencies captured by the underlying model and cannot account for instrumental effects or foreground components that are not included in the training set. The identified trends may consequently evolve when additional instrumental effects, more complex foreground models, or different prior assumptions are introduced.

The trained models provide a fast approximation of the mapping between instrumental parameters and scientific performance. The proposed ML approach thus provides a useful and efficient framework for exploring high-dimensional instrumental parameter spaces and hence
identifying instrumental optimisation directions, limited mostly by the fidelity of instrument and foreground models required to assess the actual feasibility of the experimental configurations.

\subsection{Sensitivity versus frequency coverage}
\label{subsec:sensitivity_frequency_coverage}

The results presented in Sect.~\ref{subsubsec:gb_results} illustrate a fundamental trade-off between instrumental sensitivity and frequency coverage. The optimal instrument configuration depends not only on the overall sensitivity but also on how this sensitivity is distributed across the frequency range. This raises the question of how the choice of frequency coverage affects the ability to recover the target signals and whether extending the observing band is always beneficial.

Extending the observing band provides additional spectral information that can help characterise and separate foreground components. However, this necessarily reduces the sensitivity available at a given frequency. The optimal balance between these two effects therefore depends on which aspect of the measurement is limiting the recovery of the target signal. This distinction is particularly important when considering the complexity of the foreground model. 

For a relatively simple foreground model or when the foreground properties are already well constrained by external data, the additional information provided by a broader frequency range may have a limited impact on the final constraints. In this regime, concentrating the available instrumental resources over the most relevant frequency range can be advantageous, as improving the sensitivity directly enhances the measurement of the target signal. Conversely, when several foreground components with partially degenerate spectral signatures must be simultaneously constrained, frequency coverage becomes an essential source of information. Additional spectral channels can help to distinguish the different components and reduce degeneracies with the CMB spectral distortions. In this regime, sacrificing some sensitivity per frequency channel in favour of broader frequency coverage can result in better overall constraints on the target signals.

The CIB case provides a good example of this latter regime (see Fig. \ref{fig: SHAP cib}). The results from both direct and ML optimisations indicate that the instrument configuration should not necessarily maximise the raw sensitivity. Instead, it should balance sensitivity against spectral coverage depending on the complexity of the sky model and the available prior information on the foregrounds. As the foreground model becomes increasingly complex or poorly constrained, the importance of broad frequency coverage increases, potentially favouring an instrument with lower sensitivity per frequency channel but greater spectral leverage (see Fig.\ref{fig:f_min_trend}). 

The analysis performed here does not make use of analysis spatial information. In a more realistic analysis, the choice of scanning strategy and the resulting spatial coverage could provide additional information to separate the target signal from the foregrounds, particularly since foreground components exhibit distinct spatial morphologies. This information could modify the relative importance of sensitivity and frequency coverage, providing additional means of constraining foregrounds and consequently altering the optimal allocation of instrumental resources. Nevertheless, it would not change the fundamental trade-off between these two aspects: broader frequency coverage would still provide additional spectral leverage, while distributing the available instrumental resources over a wider band would generally reduce the sensitivity achieved at individual frequencies. 

\subsection{Limitations of the optimisation framework}
\label{subsec: limitations}

The present analysis relies on a simplified instrument model, which has several consequences. Most notably, the adopted model reduces the thermal emission budget to the warmest optical component only. This approximation is sufficient for a first-order sensitivity analysis but it underestimates the complexity of internal emission contributions in a realistic instrument. In addition, certain parameters of the instrument were not taken into account. For example, the number of detectors is not explicitly included in the model. It is rather based on an equivalence at the level of photon noise scaling. However, this simplification does not fully capture the optical couplings and the interplay between optical design, beam size, or focal plane architecture with the detectors.

The sky model itself introduces additional limitations. The emission templates used in this work are assumed to perfectly match the simulated sky. In particular, thermal dust emission is approximated by a modified blackbody, whereas more complex models, such as multi-component dust descriptions, may significantly alter spectral degeneracies at high frequencies \citep{hensley2022astrodust, ysard2024themis2}. Such effects become increasingly relevant above the \textit{Planck} frequency range (857~GHz), especially for future missions extending beyond $\sim 5$~THz \citep{Planck2016Xforeground_planck_maps}. In that regime, deviations from simple grey-body assumptions may alter the inferred relative importance of high-frequency coverage in the instrument optimisation.

On the methodology part, the ML approach we use for the instrument optimisation evaluates instrumental performance by exploring a high-dimensional parameter space under the assumption of independent parameter variations. While this assumption is acceptable for a simplified instrument model, it does not fully reflect the strong physical and engineering relations that exist in real experimental designs. In practice, parameters such as frequency coverage, optical throughput, detector count, NEP, and beam configuration are jointly constrained by system-level trade-offs, including optical design, cryogenic cooling capacity, and mission architecture. As a consequence, the parameter space obtained with the ML should be interpreted as an effective design rather than a physically achievable set of configurations.

In addition, the results are intrinsically dependent on the choice of parameter ranges. In particular, the relative importance of frequency bounds is strongly sensitive to the explored interval. For example, restricting $\nu_{\min}$ to values below the characteristic frequencies of the $y$-distortion leads to a reduced apparent impact of this parameter, whereas extending the band to higher frequencies significantly modifies its inferred importance. A similar effect applies to $\nu_{\max}$, especially for the CIB monopole, whose spectral peak lies in the high-frequency regime. 

Finally, the framework does not include sky-coverage optimisation, which is assumed to be fixed. However, the observing strategy plays an important role in spectral distortion measurements, as it determines both foreground contamination and statistical weight across the sky. A full optimisation would require a time-dependent pointing strategy $\vec{p}(t)$ defined over the full mission duration, which is beyond the scope of the present study and is addressed in dedicated mission-level analyses such as those of FOSSIL \citep{Coulon2026FOSSIL_SKY} and BISOU \citep{Coulon2027BISOU}.


\section{Conclusion}
\label{subsec: conclu}

In this work, we have used ML methods for the global exploration and optimisation of complex astronomical instrument designs. The approach learns the non-linear mapping between instrumental parameters and the resulting scientific performance, providing an approximation of forecasted measurement capabilities over a multidimensional instrumental parameter space. Once trained, the models can be used to identify the parameters that most strongly drive the predicted constraints and to explore regions of the parameter space that are most favourable for the scientific objectives, without requiring an exhaustive evaluation of all possible instrument configurations.

We implemented the optimisation method using decision-tree-based regression algorithms. We showed that Random Forests provide a robust first assessment of the global parameter dependencies, while gradient boosting method provides a more accurate approximation of the underlying mapping. For the three observables considered in our analysis, gradient boosting consistently outperforms Random Forests. The combination of \texttt{XGBoost} with SHAP values further provides us with a way to quantify and interpret the relative importance of the instrumental parameters. The consistency between the different diagnostics and the two ML approaches supports the robustness of the identified trends, while highlighting the importance of assessing their physical consistency rather than relying solely on predictive performance.

We applied this new ML framework to the optimisation of a wide-band Martin--Puplett FTS for CMB spectral distortion measurements, using the FOSSIL mission concept as a case study. The analysis considers the forecasted constraints on the $y$- and $\mu$-distortion monopoles and on the CIB monopole amplitude, based on a spectro-photometric model combining the relevant instrumental and sky contributions. For all three observables, we find that the temperature of the warmest instrumental component is consistently identified as the most influential parameter. This result emphasises the importance of controlling internal instrumental emission when targeting faint CMB spectral distortion signals. It supports the design of a fully cryogenic instrument, with optical and structural components maintained at temperatures of the order of $4.5\,\mathrm{K}$, wherever technically feasible \citep{Sauvage2026FOSSIL_thermal}.

Frequency coverage is also consistently identified as one of the dominant instrumental parameters, but its impact depends on the observable and on the assumed foreground model. The optimal frequency coverage is consequently not universal and depends on the target observable and on the level of foreground complexity that must be simultaneously constrained. This behaviour reflects a fundamental trade-off between spectral coverage and instrumental sensitivity. For the $\mu$-distortion, the gain from improved foreground characterisation with a broad frequency coverage can outweigh the associated loss in sensitivity, since a significant fraction of its constraining power comes from the low-frequency part of the spectrum. In contrast, the $y$-distortion is approximately two orders of magnitude larger and has a spectral signature that peaks at higher frequencies. Its measurement is therefore less dependent on the detailed characterisation of low-frequency foregrounds and more strongly driven by the available instrumental sensitivity. 

The results presented here demonstrate the potential of ML methods as an intermediate layer between physical forecasting and detailed end-to-end instrument simulations. Compared with one-parameter-at-a-time approaches, the proposed optimisation method provides a more efficient way to explore multidimensional design spaces and to identify parameter combinations that warrant further investigation. It is not intended to replace the underlying physical instrument model or more detailed simulations but rather to accelerate the exploration of their parameter space and guide subsequent higher-fidelity studies.

This work hence provides a proof of concept for machine-learning-assisted optimisation of astronomical instruments. To the best of our knowledge, it is the first application of machine-learning methods to the global optimisation of an astronomical instrument. The framework can be readily adapted to other instrument concepts and scientific objectives, provided that an appropriate physical forecasting model is available. In the present case, future developments could extend the parameter space to include additional instrumental and mission-level effects, incorporate more complex foreground models and priors, and assess how the identified optimisation trends evolve as increasingly realistic instrument models and end-to-end simulations are introduced.

\begin{acknowledgements}
The authors acknowledge financial support from the Centre national d’´etudes spatiales (CNES), France (ROR: https://ror.org/04h1h0y33).  XC knowledge financial support from  France Travail. Some of the results have been derived using the \texttt{astropy} \citep{Astropy2018} , \texttt{healpy}/\texttt{HEALPIX} \citep{Zonca2019_healpy,Gorski2005_HEALPix}, \texttt{NumPy} \citep{Harris2020_NumPy}, \texttt{SciPy} \citep{Virtanen2020_SciPy} and \texttt{Matplotlib} libraries \citep{Hunter2007_Matplotlib}.

\end{acknowledgements}

%

\bibliographystyle{aa}
\bibliography{bibliography}

@ARTICLE{Cyr2026SciencePaper,
    author = {{Cyr}, Bryce and Aghanim, Nabila and  others},
    title = "{A Science case for CMB Spectral Distortions}",
    journal = {in prep.},
    year = 2026,
}

@article{LeGall2026FOSSIL_optics,
    author = {{Loquet Le Gall}, Morgane and others},
    title = "{Optical architecture}",
    journal = {in prep.}, 
    archivePrefix = {arXiv},
    eprint = {},
    primaryClass = {astro-ph.CO},
    year = 2026,
    month = nov
}

@article{Coulon2027BISOU,
    author = {{Coulon}, Xavier and others},
    title = "{BISOU forecasts}",
    journal = {in prep.},
    archivePrefix = {arXiv},
    eprint = {},
    primaryClass = {astro-ph.CO},
    year = 2026,
    month = dec
}

@article{Coulon2026FOSSIL_SKY,
    author = {{Coulon}, Xavier and others},
    title = "{FOSSIL forecasts}",
    journal = {in prep.},
    archivePrefix = {arXiv},
    eprint = {},
    primaryClass = {astro-ph.CO},
    year = 2026,
    month = oct
}

@ARTICLE{Sauvage2026FOSSIL_thermal,
       author = {{Sauvage}, Valentin and {de Jabrun}, Cl{\'e}mence and {Besnard}, Ana{\"\i}s and {Borgo}, Bruno and {Charles}, Ivan and {Duval}, Jean-Marc and {Maffei}, Bruno and {Martin}, Sylvain and {Aghanim}, Nabila},
        title = "{FOSSIL's preliminary thermal architecture}",
      journal = {arXiv e-prints},
         year = 2026,
        month = aug,
          eid = {arXiv:2608.13185},
        pages = {arXiv:2608.13185},
          doi = {10.48550/arXiv.2608.13185},
archivePrefix = {arXiv},
       eprint = {2608.13185},
 primaryClass = {astro-ph.IM},
       adsurl = {https://ui.adsabs.harvard.edu/abs/2026arXiv260813185S}
}

@article{Aghanim2026FOSSIL_general_paper,
       author = {{FOSSIL Collaboration} and Aghanim, N. and Maffei, B. and Aumont, J. and Beelen, A. and Borgo, B. and Bozzo, E. and Chluba, J. and Coulon, X. and Couturier, S. and Cuttaia, F. and Cyr, B. and De Bernardis, P. and De Jabrun, C. and Díaz-García, J.-J. and Eckert, D. and Fabbian, G. and Ferrari, L. and Finelli, F. and Gascard, T. and Gudmundsson, J. E. and Kogut, A. and Lagache, G. and Loquet Le Gall, M. and Martin, S. and Martins, A. I. S. and Monfardini, A. and O'Sullivan, C. and Pagano, L. and Paoletti, D. and Pisano, G. and Remazeilles, M. and Reveret, V. and Rubino-Martin, J. A. and Sauvage, V. and Savini, G. and Spencer, L. and Tartari, A. and Terenzi, L. and Trappe, N. and Watts, D. and Winter, B. and Baker, E. and Battistelli, E. S. and Battye, R. and Bernal, J. L. and Besnard, A. and Bouchet, F. R. and Braglia, M. and Burigana, C. and Calvo, M. and Caminade, S. and Citran, M. and Clements, D. and Coulton, W. and Courau, E. and De Zotti, G. and Di Valentino, E. and Dole, H. and Domenech, G. and Evangelista, S. and Farrens, S. and Frailis, M. and Glasscock, K. A. and Hervier, V. and Kovetz, E. and Liu, H. and Louis, T. and Madec, F. and Maniyar, A. S. and Martins, C. J. A. P. and Masi, S. and Maurin, L. and Moll, F. and Morelli, L. and Naess, S. K. and Nicola, A. and Pace, F. and Ponthieu, N. and Poulin, V. and Prêle, E. and Sanchez, N. and Qin, W. and Rodriguez, L. and Scott, D. and Simon, F. and Singal, J. and Slatyer, T. and Teixeira, E. M. and Thiele, L. and Tristram, M. and Trombetti, T. and Vacher, L. and Vinzl, S. and Zacchei, A. and Allys, E. and Aryan, K. and Casas Gonzales, J. M. and Lahav, O. and Sidiropoulos, C. and others},
        title = "{FOSSIL: A future mission for CMB spectral distortion measurement}",
      journal = {arXiv e-prints},
         year = "2026",
        month = "Sept",
 archivePrefix = {arXiv},
       eprint = {2608.xxxxx},
 primaryClass = {astro-ph.CO}
}

@ARTICLE{DeZotti2010_radio_sources,
       author = {{De Zotti}, Gianfranco and {Massardi}, Marcella and {Negrello}, Mattia and {Wall}, Jasper},
        title = "{Radio and millimeter continuum surveys and their astrophysical implications}",
      journal = {\aapr},
         year = 2010,
        month = feb,
       volume = {18},
       number = {1-2},
        pages = {1-65},
          doi = {10.1007/s00159-009-0026-0},
archivePrefix = {arXiv},
       eprint = {0908.1896},
 primaryClass = {astro-ph.CO},
       adsurl = {https://ui.adsabs.harvard.edu/abs/2010A&ARv..18....1D}
}

@ARTICLE{Zonca2019_healpy,
       author = {{Zonca}, Andrea and {Singer}, Leo P. and {Lenz}, Daniel and
                 {Reinecke}, Martin and {Rosset}, Cyrille and {Hivon}, Eric and
                 {Gorski}, Krzysztof M.},
        title = "{healpy: equal area pixelization and spherical harmonics transforms for data on the sphere in Python}",
      journal = {Journal of Open Source Software},
         year = 2019,
       volume = {4},
       number = {35},
        pages = {1298},
          doi = {10.21105/joss.01298}
}

@INPROCEEDINGS{Mather1993FIRAS_MTM,
       author = {{Mather}, John C. and {Fixsen}, Dale J. and {Shafer}, Richard A.},
        title = "{Design for the COBE far-infrared absolute spectrophotometer (FIRAS)}",
    booktitle = {Infrared Spaceborne Remote Sensing},
         year = 1993,
       editor = {{Scholl}, Marija S.},
       series = {Society of Photo-Optical Instrumentation Engineers (SPIE) Conference Series},
       volume = {2019},
        month = oct,
        pages = {168-179},
          doi = {10.1117/12.157823},
       adsurl = {https://ui.adsabs.harvard.edu/abs/1993SPIE.2019..168M}
}

@inproceedings{Stark1986_MTM,
  author    = {Stark, Kenneth W. and Wilson, Meredith},
  title     = {A Mirror Transport Mechanism for Use at Cryogenic Temperatures},
  booktitle = {20th Aerospace Mechanisms Symposium},
  year      = {1986},
  month     = may
}

@ARTICLE{Gorski2005_HEALPix,
       author = {{G{\'o}rski}, Krzysztof M. and {Hivon}, Eric and
                 {Banday}, A. J. and {Wandelt}, Benjamin D. and
                 {Hansen}, Frode K. and {Reinecke}, Martin and
                 {Bartelmann}, Matthias},
        title = "{HEALPix: A Framework for High-Resolution Discretization and Fast Analysis of Data Distributed on the Sphere}",
      journal = {\apj},
         year = 2005,
       month = apr,
       volume = {622},
       number = {2},
        pages = {759--771},
          doi = {10.1086/427976},
archivePrefix = {arXiv},
       eprint = {astro-ph/0409513},
 primaryClass = {astro-ph}
}

@ARTICLE{Harris2020_NumPy,
       author = {{Harris}, Charles R. and {Millman}, K. Jarrod and
                 {van der Walt}, St{\'e}fan J. and {Gommers}, Ralf and
                 {Virtanen}, Pauli and {Cournapeau}, David and {Wieser}, Eric and
                 {Taylor}, Julian and {Berg}, Sebastian and {Smith}, Nathaniel J. and
                 {Kern}, Robert and {Picus}, Matti and {Hoyer}, Stephan and
                 {van Kerkwijk}, Marten H. and {Brett}, Matthew and
                 {Haldane}, Allan and {Fern{\'a}ndez del R{\'i}o}, Jaime and
                 {Wiebe}, Mark and {Peterson}, Pearu and
                 {G{\'e}rard-Marchant}, Pierre and {Sheppard}, Kevin and
                 {Reddy}, Tyler and {Weckesser}, Warren and {Abbasi}, Hameer and
                 {Gohlke}, Christoph and {Oliphant}, Travis E.},
        title = "{Array programming with NumPy}",
      journal = {Nature},
         year = 2020,
       month = sep,
       volume = {585},
       number = {7825},
        pages = {357--362},
          doi = {10.1038/s41586-020-2649-2},
archivePrefix = {arXiv},
       eprint = {2006.10256},
 primaryClass = {cs.MS}
}

@article{Catalano2020_KIDs,
  author  = {Catalano, A. and others},
  title   = {Sensitivity of {LEKID} for space applications between 80~{GHz} and 600~{GHz}},
  journal = {Astron. Astrophys.},
  year    = {2020},
  volume  = {641},
  pages   = {A179},
  doi     = {10.1051/0004-6361/202038199}
}

@ARTICLE{LeGall2026optics_BISOU_breadboard,
       author = {{Loquet Le Gall}, Morgane and {O'Sullivan}, Creidhe and {Borgo}, Bruno and {De Jabrun}, Cl{\'e}mence and {Sauvage}, Valentin and {Trappe}, Neil and {Maffei}, Bruno},
        title = "{Optical development of the BISOU breadboard}",
      journal = {arXiv e-prints},
         year = 2026,
        month = aug,
          eid = {arXiv:2608.13225},
        pages = {arXiv:2608.13225},
          doi = {10.48550/arXiv.2608.13225},
archivePrefix = {arXiv},
       eprint = {2608.13225},
 primaryClass = {astro-ph.IM},
       adsurl = {https://ui.adsabs.harvard.edu/abs/2026arXiv260813225L}
}

@ARTICLE{Virtanen2020_SciPy,
       author = {{Virtanen}, Pauli and {Gommers}, Ralf and
                 {Oliphant}, Travis E. and {Haberland}, Matt and
                 {Reddy}, Tyler and {Cournapeau}, David and {Burovski}, Evgeni and
                 {Peterson}, Pearu and {Weckesser}, Warren and {Bright}, Jonathan and
                 {van der Walt}, St{\'e}fan J. and {Brett}, Matthew and
                 {Wilson}, Joshua and {Millman}, K. Jarrod and {Mayorov}, Nikolay and
                 {Nelson}, Andrew R.~J. and {Jones}, Eric and {Kern}, Robert and
                 {Larson}, Eric and {Carey}, C.~J. and {Polat}, {\.I}lhan and
                 {Feng}, Yu and {Moore}, Eric W. and {VanderPlas}, Jake and
                 {Laxalde}, Denis and {Perktold}, Josef and {Cimrman}, Robert and
                 {Henriksen}, Ian and {Quintero}, E.~A. and {Harris}, Charles R. and
                 {Archibald}, Anne M. and {Ribeiro}, Ant{\^o}nio H. and
                 {Pedregosa}, Fabian and {van Mulbregt}, Paul and
                 {{SciPy 1.0 Contributors}}},
        title = "{SciPy 1.0: Fundamental Algorithms for Scientific Computing in Python}",
      journal = {Nature Methods},
         year = 2020,
       month = mar,
       volume = {17},
       number = {3},
        pages = {261--272},
          doi = {10.1038/s41592-019-0686-2},
archivePrefix = {arXiv},
       eprint = {1907.10121},
 primaryClass = {cs.MS}
}

@ARTICLE{Astropy2018,
author = {{Astropy Collaboration} and {Price-Whelan}, Adrian M. and
{Sip{\H{o}}cz}, Brigitta M. and
{G{"u}nther}, H. Moritz and
{Lim}, Pey Lian and
{Crawford}, Steven M. and
{Conseil}, Simon and
{Shupe}, David L. and
{Craig}, Matthew W. and
{Dencheva}, Nadia and
{Ginsburg}, Adam and
{VanderPlas}, Jacob T. and
{Bradley}, Larry D. and
{P{'e}rez-Su{'a}rez}, David and
{de Val-Borro}, Miguel and
{{Paper Contributors}} and
{Aldcroft}, Thomas L. and
{Cruz}, Kelle L. and
{Robitaille}, Thomas P. and
{Tollerud}, Erik J. and
{{Coordination Committee}} and
{Kirkby}, David and
{Mizuno}, Matthew and
{Bach}, Yurii and
{Bachetti}, Matteo and others},
journal = {\aj},
year = 2018,
month = sep,
volume = {156},
number = {3},
eid = {123},
pages = {123},
doi = {10.3847/1538-3881/aabc4f},
archivePrefix = {arXiv},
eprint = {1801.02634},
primaryClass = {astro-ph.IM},
adsurl = {https://ui.adsabs.harvard.edu/abs/2018AJ....156..123A}
}

@ARTICLE{Hunter2007_Matplotlib,
       author = {{Hunter}, John D.},
        title = "{Matplotlib: A 2D Graphics Environment}",
      journal = {Computing in Science \& Engineering},
         year = 2007,
       volume = {9},
       number = {3},
        pages = {90--95},
          doi = {10.1109/MCSE.2007.55}
}

@article{Baselmans2022kids,
       author = {{Baselmans}, J.~J.~A. and {Facchin}, F. and {Pascual Laguna}, A. and {Bueno}, J. and {Thoen}, D.~J. and {Murugesan}, V. and {Llombart}, N. and {de Visser}, P.~J.},
        title = "{Ultra-sensitive THz microwave kinetic inductance detectors for future space telescopes}",
      journal = {\aap},
         year = 2022,
        month = sep,
       volume = {665},
          eid = {A17},
        pages = {A17},
          doi = {10.1051/0004-6361/202243840},
archivePrefix = {arXiv},
       eprint = {2207.08647},
 primaryClass = {astro-ph.IM},
       adsurl = {https://ui.adsabs.harvard.edu/abs/2022A&A...665A..17B}
}

@INPROCEEDINGS{Landman2020_ML_opti_adaptive_optics_exemple,
    author    = {{Landman}, R. and {Haffert}, S. Y. and
                 {Radhakrishnan}, V. M. and {Keller}, C. U.},
    title     = {Self-optimizing adaptive optics control with reinforcement learning},
    booktitle = {Adaptive Optics Systems VII},
    year      = 2020,
    editor    = {{Schmidt}, D. and {Schreiber}, L. and {Vernet}, E.},
    series    = {Society of Photo-Optical Instrumentation Engineers (SPIE) Conference Series},
    volume    = 11448,
    month     = dec,
    eid       = {114480K},
    doi       = {10.1117/12.2560053}
}

@ARTICLE{Pou2022_ML_opti_adaptive_optics_exemple,
    author  = {{Pou}, B. and {Ferreira}, F. and {Quinones}, E. and
               {Gratadour}, D. and {Martin}, M.},
    title   = {Adaptive optics control with multi-agent model-free reinforcement learning},
    journal = {Optics Express},
    year    = 2022,
    month   = jan,
    volume  = 30,
    number  = 2,
    pages   = {2991--3015},
    doi     = {10.1364/OE.444099}
}

@ARTICLE{Qasim2024_ML_opti_detector_design_exemple,
    author  = {{Qasim}, Shah Rukh and {Owen}, Patrick and {Serra}, Nicola},
    title   = {Physics instrument design with reinforcement learning},
    journal = {Machine Learning: Science and Technology},
    year    = 2025,
    volume  = 6,
    number  = 3,
    pages   = {035003},
    doi     = {10.1088/2632-2153/adf7ff},
    archivePrefix = {arXiv},
    eprint  = {2412.10237},
    primaryClass = {physics.ins-det}
}

@inproceedings{Manzan2024_CMBspecDist_COSMO,
  author = {Manzan, E. and Albano, L. and Franceschet, C. and Battistelli, E. S. and de Bernardis, P. and Bersanelli, M. and Cacciotti, F. and Capponi, A. and Columbro, F. and Conenna, G. and Coppi, G. and Coppolecchia, A. and D’Alessandro, G. and De Gasperis, G. and De Petris, M. and Gervasi, M. and Isopi, G. and Lamagna, L. and Limonta, A. and Marchitelli, E. and Masi, S. and Mennella, A. and Montonati, F. and Nati, F. and Occhiuzzi, A. and Paiella, A. and Pettinari, G. and Piacentini, F. and Piccirillo, L. and Pisano, G. and Tucker, C. and Zannoni, M.},
  title = {Measuring the CMB spectral distortions with COSMO: the multi‑mode antenna system},
  booktitle = {Millimeter, Submillimeter, and Far‑Infrared Detectors and Instrumentation for Astronomy XII},
  series = {Proceedings of SPIE},
  volume = {13102},
  pages = {131021A},
  year = {2024},
  doi = {10.1117/12.3018730},
  url = {https://arxiv.org/abs/2406.09349}
}

@article{Mather1982_BolometerNoise,
  author = {Mather, J. C.},
  title = {Bolometer Noise: Nonequilibrium Theory},
  journal = {Applied Optics},
  volume = {21},
  number = {6},
  pages = {1125--1129},
  year = {1982},
  doi = {10.1364/AO.21.001125},
  url = {https://opg.optica.org/ao/abstract.cfm?uri=ao-21-6-1125}
}

@article{Lamarre1989,
  author  = {Lamarre, Jean-Michel},
  title   = {Photon noise in photometric instruments at far-infrared and submillimeter wavelengths},
  journal = {Applied Optics},
  volume  = {25},
  number  = {6},
  pages   = {870--876},
  year    = {1986},
  doi     = {10.1364/AO.25.000870}
}

@misc{PySM3,
  author = {Zonca, Andrea and Thorne, Ben and Krachmalnicoff, Nicoletta and Borrill, Julian},
  title = {PySM 3: The Python Sky Model 3 software},
  year = {2021},
  url = {https://pysm3.readthedocs.io/en/latest/}
}

@inproceedings{Mather1993firas,
  author = {Mather, J. C. and Fixsen, D. J. and Shafer, R. A.},
  title = {Design for the COBE Far Infrared Absolute Spectrophotometer (FIRAS)},
  booktitle = {Proceedings of SPIE, Infrared Spaceborne Remote Sensing},
  volume = {2019},
  pages = {168--179},
  year = {1993},
  doi = {10.1117/12.157823}
}

@ARTICLE{CTA2019_opti_CTA,
       author = {{Acharyya}, A. and {Agudo}, I. and {Ang{\"u}ner}, E.~O. and {Alfaro}, R. and {Alfaro}, J. and {Alispach}, C. and {Aloisio}, R. and {Alves Batista}, R. and {Amans}, J.-P. and {Amati}, L. and {Amato}, E. and {Ambrosi}, G. and {Antonelli}, L.~A. and {Aramo}, C. and {Armstrong}, T. and {Arqueros}, F. and {Arrabito}, L. and {Asano}, K. and {Ashkar}, H. and {Balazs}, C. and {Balbo}, M. and {Balmaverde}, B. and {Barai}, P. and {Barbano}, A. and {Barkov}, M. and {Barres de Almeida}, U. and {Barrio}, J.~A. and {Bastieri}, D. and {Becerra Gonz{\'a}lez}, J. and {Becker Tjus}, J. and {Bellizzi}, L. and {Benbow}, W. and {Bernardini}, E. and {Bernardos}, M.~I. and {Bernl{\"o}hr}, K. and {Berti}, A. and {Berton}, M. and {Bertucci}, B. and {Beshley}, V. and {Biasuzzi}, B. and {Bigongiari}, C. and {Bird}, R. and {Bissaldi}, E. and {Biteau}, J. and {Blanch}, O. and {Blazek}, J. and {Boisson}, C. and {Bonanno}, G. and {Bonardi}, A. and {Bonavolont{\'a}}, C. and {Bonnoli}, G. and {Bordas}, P. and {B{\"o}ttcher}, M. and {Bregeon}, J. and {Brill}, A. and {Brown}, A.~M. and {Br{\"u}gge}, K. and {Brun}, P. and {Bruno}, P. and {Bulgarelli}, A. and {Bulik}, T. and {Burton}, M. and {Burtovoi}, A. and {Busetto}, G. and {Cameron}, R. and {Canestrari}, R. and {Capalbi}, M. and {Caproni}, A. and {Capuzzo-Dolcetta}, R. and {Caraveo}, P. and {Caroff}, S. and {Carosi}, R. and {Casanova}, S. and {Cascone}, E. and {Cassol}, F. and {Catalani}, F. and {Catalano}, O. and {Cauz}, D. and {Cerruti}, M. and {Chaty}, S. and {Chen}, A. and {Chernyakova}, M. and {Chiaro}, G. and {Cie{\'s}lar}, M. and {Colak}, S.~M. and {Conforti}, V. and {Congiu}, E. and {Contreras}, J.~L. and {Cortina}, J. and {Costa}, A. and {Costantini}, H. and {Cotter}, G. and {Cristofari}, P. and {Cumani}, P. and {Cusumano}, G. and {D'A{\'\i}}, A. and {D'Ammando}, F. and {Dangeon}, L. and {Da Vela}, P. and {Dazzi}, F. and {De Angelis}, A. and {De Caprio}, V. and {de C{\'a}ssia dos Anjos}, R. and {De Frondat}, F. and {de Gouveia Dal Pino}, E.~M. and {De Lotto}, B. and {De Martino}, D. and {de Naurois}, M. and {de O{\~n}a Wilhelmi}, E. and {de Palma}, F. and {de Souza}, V. and {Del Santo}, M. and {Delgado}, C. and {della Volpe}, D. and {Di Girolamo}, T. and {Di Pierro}, F. and {Di Venere}, L. and {D{\'\i}az}, C. and {Diebold}, S. and {Djannati-Ata{\"\i}}, A. and {Dmytriiev}, A. and {Dominis Prester}, D. and {Donini}, A. and {Dorner}, D. and {Doro}, M. and {Dournaux}, J.-L. and {Ebr}, J. and {Ekoume}, T.~R.~N. and {Els{\"a}sser}, D. and {Emery}, G. and {Falceta-Goncalves}, D. and {Fedorova}, E. and {Fegan}, S. and {Feng}, Q. and {Ferrand}, G. and {Fiandrini}, E. and {Fiasson}, A. and {Filipovic}, M. and {Fioretti}, V. and {Fiori}, M. and {Flis}, S. and {Fonseca}, M.~V. and {Fontaine}, G. and {Freixas Coromina}, L. and {Fukami}, S. and {Fukui}, Y. and {Funk}, S. and {F{\"u}{\ss}ling}, M. and {Gaggero}, D. and {Galanti}, G. and {Garcia L{\'o}pez}, R.~J. and {Garczarczyk}, M. and {Gascon}, D. and {Gasparetto}, T. and {Gaug}, M. and {Ghalumyan}, A. and {Gianotti}, F. and {Giavitto}, G. and {Giglietto}, N. and {Giordano}, F. and {Giroletti}, M. and {Gironnet}, J. and {Glicenstein}, J.-F. and {Gnatyk}, R. and {Goldoni}, P. and {Gonz{\'a}lez}, J.~M. and {Gonz{\'a}lez}, M.~M. and {Gourgouliatos}, K.~N. and {Grabarczyk}, T. and {Granot}, J. and {Green}, D. and {Greenshaw}, T. and {Grondin}, M.-H. and {Gueta}, O. and {Hadasch}, D. and {Hassan}, T. and {Hayashida}, M. and {Heller}, M. and {Hervet}, O. and {Hinton}, J. and {Hiroshima}, N. and {Hnatyk}, B. and {Hofmann}, W. and {Horvath}, P. and {Hrabovsky}, M. and {Hrupec}, D. and {Humensky}, T.~B. and {H{\"u}tten}, M. and {Inada}, T. and {Iocco}, F. and {Ionica}, M. and {Iori}, M. and {Iwamura}, Y. and {Jamrozy}, M. and {Janecek}, P. and {Jankowsky}, D. and {Jean}, P. and {Jouvin}, L. and {Jurysek}, J. and {Kaaret}, P.},
        title = "{Monte Carlo studies for the optimisation of the Cherenkov Telescope Array layout}",
      journal = {Astroparticle Physics},
         year = 2019,
        month = sep,
       volume = {111},
        pages = {35-53},
          doi = {10.1016/j.astropartphys.2019.04.001},
archivePrefix = {arXiv},
       eprint = {1904.01426},
 primaryClass = {astro-ph.IM},
       adsurl = {https://ui.adsabs.harvard.edu/abs/2019APh...111...35A}
}

@ARTICLE{Prince2002_opti_LISA,
       author = {{Prince}, Thomas A. and {Tinto}, Massimo and {Larson}, Shane L. and {Armstrong}, J.~W.},
        title = "{LISA optimal sensitivity}",
      journal = {\prd},
         year = 2002,
        month = dec,
       volume = {66},
       number = {12},
          eid = {122002},
        pages = {122002},
          doi = {10.1103/PhysRevD.66.122002},
archivePrefix = {arXiv},
       eprint = {gr-qc/0209039},
 primaryClass = {gr-qc},
       adsurl = {https://ui.adsabs.harvard.edu/abs/2002PhRvD..66l2002P}
}

@article{Dorigo2023_opti_AD_vs_ML,
  title     = {Toward the end-to-end optimization of particle physics instruments with differentiable programming},
  author    = {Dorigo, Tommaso and Giammanco, Andrea and Vischia, Pietro and Aehle, Max and Bawaj, Mateusz and Boldyrev, Alexey and de Castro Manzano, Pablo and Derkach, Denis and Donini, Julien and Edelen, Auralee and Fanzago, Federica and Gauger, Nicolas R. and Glaser, Christian and Baydin, At{\i}l{\i}m G. and Heinrich, Lukas and Keidel, Ralf and Kieseler, Jan and Krause, Claudius and Lagrange, Maxime and Lamparth, Max and Layer, Lukas and Maier, Gernot and Nardi, Federico and Pettersen, Helge E. S. and Ramos, Alberto and Ratnikov, Fedor and R{\"o}hrich, Dieter and Ruiz de Austri, Roberto and Ruiz del {\'A}rbol, Pablo Mart{\'\i}nez and Savchenko, Oleg and Simpson, Nathan and Strong, Giles C. and Taliercio, Angela and Tosi, Mia and Ustyuzhanin, Andrey and Zaraket, Haitham},
  journal   = {Reviews in Physics},
  volume    = {10},
  pages     = {100085},
  year      = {2023},
  publisher = {Elsevier},
  doi       = {10.1016/j.revip.2023.100085},
  issn      = {2405-4283}
}

@article{Arectout2021_opti_det,
  title     = {Optimization of the n-type HPGe detector parameters using the ``design of experiments'' technique},
  author    = {Arectout, Assia and Boukhal, H. and Chakir, E. and Chham, Essaid and Ferro-Garc{\'\i}a, M. A. and Pi{\~n}ero-Garc{\'\i}a, F. and Azahra, M. and Makhloul, M. and Azougagh, M. and El Yaakoubi, H. and Zidouh, I. and Exp{\'o}sito-Su{\'a}rez, V. M.},
  journal   = {Radiation Physics and Chemistry},
  volume    = {189},
  pages     = {109733},
  year      = {2021},
  month     = {December},
  publisher = {Elsevier},
  doi       = {10.1016/j.radphyschem.2021.109733},
  issn      = {0969-806X}
}

@article{Knapp2023_opti_plasma,
  title     = {Optimizing the configuration of plasma radiation detectors in the presence of uncertain instrument response and inadequate physics},
  author    = {Knapp, Patrick F. and Lewis, William E. and Joseph, V. Roshan},
  journal   = {Journal of Plasma Physics},
  volume    = {89},
  number    = {1},
  pages     = {895890101},
  year      = {2023},
  publisher = {Cambridge University Press},
  doi       = {10.1017/S002237782200126X},
  issn      = {0022-3778}
}

@ARTICLE{Sethuram2023_ML_radiative_transfer_emulator_exemple,
       author = {{Sethuram}, Snigdaa S. and {Cochrane}, Rachel K. and {Hayward}, Christopher C. and {Acquaviva}, Viviana and {Villaescusa-Navarro}, Francisco and {Popping}, Gerg{\"o} and {Wise}, John H.},
        title = "{Emulating radiative transfer with artificial neural networks}",
      journal = {\mnras},
         year = 2023,
        month = dec,
       volume = {526},
       number = {3},
        pages = {4520-4528},
          doi = {10.1093/mnras/stad2524},
archivePrefix = {arXiv},
       eprint = {2308.13648},
 primaryClass = {astro-ph.GA},
       adsurl = {https://ui.adsabs.harvard.edu/abs/2023MNRAS.526.4520S}
}

@ARTICLE{Qasim2024_ML_opti_physics_instrument_design_exemple,
       author = {{Qasim}, Shah Rukh and {Owen}, Patrick and {Serra}, Nicola},
        title = "{Physics Instrument Design with Reinforcement Learning}",
      journal = {arXiv e-prints},
         year = 2024,
        month = dec,
          eid = {arXiv:2412.10237},
archivePrefix = {arXiv},
       eprint = {2412.10237},
 primaryClass = {physics.ins-det},
       adsurl = {https://ui.adsabs.harvard.edu/abs/2024arXiv241210237Q}
}

@ARTICLE{Baydin2021_ML_opti_experimental_design_exemple,
       author = {{Baydin}, Atılım G{\"u}neş and {Cranmer}, Kyle and {Sanchez-Gonzalez}, Alvaro and {Schneider}, Andy and {Dillon}, Joshua V. and {Schoenholz}, Samuel S. and {others}},
        title = "{Towards Machine Learning Optimization of Experimental Design}",
      journal = {Computing and Software for Big Science},
         year = 2021,
        volume = {5},
        pages = {1--24},
          doi = {10.1007/s41781-021-00061-7},
archivePrefix = {arXiv},
       eprint = {2106.08325},
 primaryClass = {cs.LG},
       adsurl = {https://ui.adsabs.harvard.edu/abs/2021CSBS....5....1B}
}

@ARTICLE{GranadosOrtiz2021_ML_opti_design_optimization_exemple,
       author = {{Granados-Ortiz}, Francisco-Javier and {Ortega-Casanova}, Joaqu{\'i}n},
        title = "{Machine-learning-aided design optimization of a mechanical micromixer}",
      journal = {Physics of Fluids},
       volume = {33},
       number = {6},
        pages = {063604},
         year = 2021,
          doi = {10.1063/5.0048771}
}

@ARTICLE{Mirhoseini2021_ML_opti_chip_design_exemple,
       author = {{Mirhoseini}, Azalia and {Goldie}, Anna and {Yazgan}, Mustafa and {Wang}, Joe and {Jiang}, Jiageng and {Songhori}, Eriko and {Scharf, Elad and others}},
        title = "{A Graph Placement Methodology for Fast Chip Design}",
      journal = {Nature},
       volume = {594},
        pages = {207--212},
         year = 2021,
          doi = {10.1038/s41586-021-03544-w},
archivePrefix = {arXiv},
       eprint = {2004.10746},
       adsurl = {https://ui.adsabs.harvard.edu/abs/2021Natur.594..207M}
}

@ARTICLE{Bonnaire2020_ML_filament_detection_exemple,
       author = {{Bonnaire}, Tony and {Aghanim}, Nabila and {Decelle}, Aur{\'e}lien and {Douspis}, Marian},
        title = "{T-ReX: a graph-based filament detection method}",
      journal = {\aap},
         year = 2020,
        volume = {637},
          eid = {A18},
        pages = {A18},
          doi = {10.1051/0004-6361/201936859},
archivePrefix = {arXiv},
       eprint = {2002.06723},
 primaryClass = {astro-ph.CO},
       adsurl = {https://ui.adsabs.harvard.edu/abs/2020A&A...637A..18B}
}

@ARTICLE{Mikuni2022_ML_generative_detector_simulation_exemple,
       author = {{Mikuni}, Vinicius and {Nachman}, Benjamin},
        title = "{Score-based generative models for calorimeter shower simulation}",
      journal = {\prd},
         year = 2022,
        month = nov,
       volume = {106},
       number = {9},
          eid = {092009},
        pages = {092009},
          doi = {10.1103/PhysRevD.106.092009},
archivePrefix = {arXiv},
       eprint = {2206.11898},
 primaryClass = {hep-ph},
       adsurl = {https://ui.adsabs.harvard.edu/abs/2022PhRvD.106i2009M}
}

@ARTICLE{Farid2023_ML_galaxy_cluster_classification_exemple,
       author = {{Farid}, D. and {Aung}, H. and {Nagai}, D. and {Farahi}, A. and {Rozo}, E.},
        title = "{C$^{2}$-GAME: Classification of cluster galaxy membership with machine learning}",
      journal = {Astronomy and Computing},
         year = 2023,
        month = oct,
       volume = {45},
          eid = {100743},
        pages = {100743},
          doi = {10.1016/j.ascom.2023.100743},
archivePrefix = {arXiv},
       eprint = {2205.01700},
 primaryClass = {astro-ph.CO},
       adsurl = {https://ui.adsabs.harvard.edu/abs/2023A&C....4500743F}
}

@ARTICLE{Conceicao2024_ML_cosmological_emulator_exemple,
       author = {{Concei{\c{c}}{\~a}o}, Miguel and {Krone-Martins}, Alberto and {da Silva}, Antonio and {Molin{\'e}}, {\'A}ngeles},
        title = "{Fast emulation of cosmological density fields based on dimensionality reduction and supervised machine learning}",
      journal = {\aap},
         year = 2024,
        month = jan,
       volume = {681},
          eid = {A123},
        pages = {A123},
          doi = {10.1051/0004-6361/202346734},
archivePrefix = {arXiv},
       eprint = {2304.06099},
 primaryClass = {astro-ph.CO},
       adsurl = {https://ui.adsabs.harvard.edu/abs/2024A&A...681A.123C}
}

@article{Dai2018RF,
author  = {Dai, Mi and Kuhlmann, Steve and Wang, Yun and Kovacs, Eve},
title   = {Photometric classification and redshift estimation of {LSST} Supernovae},
journal = {Monthly Notices of the Royal Astronomical Society},
volume  = {477},
number  = {3},
pages   = {4142--4151},
year    = {2018},
doi     = {10.1093/mnras/sty965}
}

@article{Richards2012RF,
author  = {Richards, Joseph W. and Homrighausen, Darren and Freeman, Peter E. and
Schafer, Chad M. and Poznanski, Dovi},
title   = {Semi-supervised learning for photometric supernova classification},
journal = {Monthly Notices of the Royal Astronomical Society},
volume  = {419},
number  = {2},
pages   = {1121--1135},
year    = {2012},
doi     = {10.1111/j.1365-2966.2011.19768.x}
}

@article{Kim2015RF,
author  = {Kim, Edward J. and Brunner, Robert J. and Carrasco Kind, Matias},
title   = {A hybrid ensemble learning approach to star--galaxy classification},
journal = {Monthly Notices of the Royal Astronomical Society},
volume  = {453},
number  = {1},
pages   = {507--521},
year    = {2015},
doi     = {10.1093/mnras/stv1608}
}

@article{ArroquiaCuadros2023RF,
author  = {Arroquia-Cuadros, Benjam{'i}n and S{'a}nchez, N{'e}stor and
G{'o}mez, Vicent and Blay, Pere and Martinez-Badenes, Vicent and
Nieves-Seoane, Lorena},
title   = {Photometric classification of quasars from {ALHAMBRA} survey using random forest},
journal = {Astronomy \& Astrophysics},
volume  = {673},
pages   = {A48},
year    = {2023},
doi     = {10.1051/0004-6361/202245531}
}

@article{FlukeJacobs2020RF,
author  = {Fluke, Christopher J. and Jacobs, Colin},
title   = {Surveying the reach and maturity of machine learning and artificial intelligence in astronomy},
journal = {WIREs Data Mining and Knowledge Discovery},
volume  = {10},
number  = {2},
pages   = {e1349},
year    = {2020},
doi     = {10.1002/widm.1349}
}

@ARTICLE{Bonjean2020_ML_SZ_detection_exemple,
       author = {{Bonjean}, V.},
        title = "{Deep learning for Sunyaev-Zel'dovich detection in Planck}",
      journal = {\aap},
         year = 2020,
        month = feb,
       volume = {634},
          eid = {A81},
        pages = {A81},
          doi = {10.1051/0004-6361/201936919},
archivePrefix = {arXiv},
       eprint = {1911.10778},
 primaryClass = {astro-ph.CO},
       adsurl = {https://ui.adsabs.harvard.edu/abs/2020A&A...634A..81B}
}

@article{Friedman2002_stochastic_boosting,
  author  = {Friedman, Jerome H.},
  title   = {Stochastic Gradient Boosting},
  journal = {Computational Statistics \& Data Analysis},
  year    = {2002},
  volume  = {38},
  number  = {4},
  pages   = {367--378},
  doi     = {10.1016/S0167-9473(01)00065-2}
}

@article{Strobl2007_BiasRFVarImp,
  author  = {Strobl, Carolin and Boulesteix, Anne-Laure and Zeileis, Achim and Hothorn, Torsten},
  title   = {Bias in Random Forest Variable Importance Measures: Illustrations, Sources and a Solution},
  journal = {BMC Bioinformatics},
  year    = {2007},
  volume  = {8},
  pages   = {25},
  doi     = {10.1186/1471-2105-8-25}
}

@ARTICLE{He2019_ML_simulation_acceleration_exemple,
       author = {{He}, Siyu and {Li}, Yipeng and {Feng}, Yu and {Ho}, Shirley and {Chen}, Xin and {Seljak}, Uro{\v{s}}},
        title = "{Learning to Optimize: A Machine Learning Approach for Accelerating Cosmological Simulations}",
      journal = {\apj},
         year = 2019,
        month = dec,
       volume = {887},
       number = {2},
          eid = {247},
        pages = {247},
          doi = {10.3847/1538-4357/ab5c4a},
archivePrefix = {arXiv},
       eprint = {1908.08215},
 primaryClass = {astro-ph.CO},
       adsurl = {https://ui.adsabs.harvard.edu/abs/2019ApJ...887..247H}
}

@ARTICLE{Wadekar2022_ML_simulation_emulator_exemple,
       author = {{Wadekar}, Dhruv and {Villaescusa-Navarro}, Francisco},
        title = "{Machine Learning Emulators for Cosmological Simulations}",
      journal = {\mnras},
         year = 2022,
        volume = {512},
       number = {1},
        pages = {1--15},
          doi = {10.1093/mnras/stac307},
archivePrefix = {arXiv},
       eprint = {2109.09770},
 primaryClass = {astro-ph.CO},
       adsurl = {https://ui.adsabs.harvard.edu/abs/2022MNRAS.512....1W}
}

@ARTICLE{Alsing2019_ML_simulation_based_inference_exemple,
       author = {{Alsing}, Justin and {Wandelt}, Benjamin},
        title = "{Generalized Bayesian Likelihood-free Inference with Neural Density Estimators}",
      journal = {\mnras},
         year = 2019,
        volume = {488},
       number = {4},
        pages = {5093--5103},
          doi = {10.1093/mnras/stz1960},
archivePrefix = {arXiv},
       eprint = {1903.00007},
 primaryClass = {astro-ph.CO},
       adsurl = {https://ui.adsabs.harvard.edu/abs/2019MNRAS.488.5093A}
}

@ARTICLE{Cranmer2020_ML_simulation_based_inference_exemple,
       author = {{Cranmer}, Miles and {Brehmer}, Johann and {Louppe}, Gilles},
        title = "{The Frontier of Simulation-Based Inference}",
      journal = {Proceedings of the National Academy of Sciences},
         year = 2020,
        volume = {117},
       number = {48},
        pages = {30055--30062},
          doi = {10.1073/pnas.1912789117},
archivePrefix = {arXiv},
       eprint = {1911.01429},
 primaryClass = {stat.ML},
       adsurl = {https://ui.adsabs.harvard.edu/abs/2020PNAS..11730055C}
}

@article{Planck2014HFIbeam_model_lim,
  author  = {{Planck Collaboration}},
  title   = {Planck 2013 results. VII. HFI time response and beams},
  journal = {Astronomy \& Astrophysics},
  volume  = {571},
  pages   = {A7},
  year    = {2014},
  doi     = {10.1051/0004-6361/201321535}
}

@techreport{Braun2019SKA_tech_report,
  author      = {Braun, Robert and Bonaldi, Anna and Bourke, Tyler and Keane, Evan and Wagg, Jeff},
  title       = {Anticipated Performance of the Square Kilometre Array -- Phase 1 (SKA1)},
  institution = {SKA Organisation},
  year        = {2019},
  eprint     = {1912.12699},
  archivePrefix = {arXiv}
}

@article{Planck2014HFIproc_model_lim,
  author  = {{Planck Collaboration}},
  title   = {Planck 2013 results. VI. High Frequency Instrument data processing},
  journal = {Astronomy \& Astrophysics},
  volume  = {571},
  pages   = {A6},
  year    = {2014},
  doi     = {10.1051/0004-6361/201321570}
}

@article{Planck2016X_compo_sep,
  author  = {{Planck Collaboration}},
  title   = {Planck 2015 results. X. Diffuse component separation: Foreground maps},
  journal = {Astronomy \& Astrophysics},
  volume  = {594},
  pages   = {A10},
  year    = {2016},
  doi     = {10.1051/0004-6361/201525967}
}

@article{abitbol_prospects_2017,
	title = {Prospects for {Measuring} {Cosmic} {Microwave} {Background} {Spectral} {Distortions} in the {Presence} of {Foregrounds}},
	volume = {471},
	issn = {0035-8711, 1365-2966},
	url = {http://arxiv.org/abs/1705.01534},
	doi = {10.1093/mnras/stx1653},journal = {Monthly Notices of the Royal Astronomical Society},
	author = {Abitbol, Maximilian H. and Chluba, Jens and Hill, J. Colin and Johnson, Bradley R.},
	month = oct,
	year = {2017},
	}

@ARTICLE{Amaro2017LISA,
       author = {{Amaro-Seoane}, Pau and {Audley}, Heather and {Babak}, Stanislav and {Baker}, John and {Barausse}, Enrico and {Bender}, Peter and {Berti}, Emanuele and {Binetruy}, Pierre and {Born}, Michael and {Bortoluzzi}, Daniele and {Camp}, Jordan and {Caprini}, Chiara and {Cardoso}, Vitor and {Colpi}, Monica and {Conklin}, John and {Cornish}, Neil and {Cutler}, Curt and {Danzmann}, Karsten and {Dolesi}, Rita and {Ferraioli}, Luigi and {Ferroni}, Valerio and {Fitzsimons}, Ewan and {Gair}, Jonathan and {Gesa Bote}, Lluis and {Giardini}, Domenico and {Gibert}, Ferran and {Grimani}, Catia and {Halloin}, Hubert and {Heinzel}, Gerhard and {Hertog}, Thomas and {Hewitson}, Martin and {Holley-Bockelmann}, Kelly and {Hollington}, Daniel and {Hueller}, Mauro and {Inchauspe}, Henri and {Jetzer}, Philippe and {Karnesis}, Nikos and {Killow}, Christian and {Klein}, Antoine and {Klipstein}, Bill and {Korsakova}, Natalia and {Larson}, Shane L and {Livas}, Jeffrey and {Lloro}, Ivan and {Man}, Nary and {Mance}, Davor and {Martino}, Joseph and {Mateos}, Ignacio and {McKenzie}, Kirk and {McWilliams}, Sean T and {Miller}, Cole and {Mueller}, Guido and {Nardini}, Germano and {Nelemans}, Gijs and {Nofrarias}, Miquel and {Petiteau}, Antoine and {Pivato}, Paolo and {Plagnol}, Eric and {Porter}, Ed and {Reiche}, Jens and {Robertson}, David and {Robertson}, Norna and {Rossi}, Elena and {Russano}, Giuliana and {Schutz}, Bernard and {Sesana}, Alberto and {Shoemaker}, David and {Slutsky}, Jacob and {Sopuerta}, Carlos F. and {Sumner}, Tim and {Tamanini}, Nicola and {Thorpe}, Ira and {Troebs}, Michael and {Vallisneri}, Michele and {Vecchio}, Alberto and {Vetrugno}, Daniele and {Vitale}, Stefano and {Volonteri}, Marta and {Wanner}, Gudrun and {Ward}, Harry and {Wass}, Peter and {Weber}, William and {Ziemer}, John and {Zweifel}, Peter},
        title = "{Laser Interferometer Space Antenna}",
      journal = {arXiv e-prints},
         year = 2017,
        month = feb,
          eid = {arXiv:1702.00786},
        pages = {arXiv:1702.00786},
          doi = {10.48550/arXiv.1702.00786},
archivePrefix = {arXiv},
       eprint = {1702.00786},
 primaryClass = {astro-ph.IM},
       adsurl = {https://ui.adsabs.harvard.edu/abs/2017arXiv170200786A}
}

@article{Laureijs2011euclid_model_instru,
  author  = {Laureijs, R. and Amiaux, J. and Arduini, S. and Augu{\`e}res, J.-L. and Brinchmann, J. and Cole, R. and Cropper, M. and Dabin, C. and Duvet, L. and Ealet, A. and others},
  title   = {Euclid Definition Study Report},
  journal = {arXiv e-prints},
  archivePrefix = {arXiv},
  eprint  = {1110.3193},
  year    = {2011}
}

@article{Tauber2010planck_model_instru,
  author  = {Tauber, J. A. and Mandolesi, N. and Puget, J.-L. and Banos, T. and Bersanelli, M. and Bouchet, F. R. and Butler, R. C. and Charra, J. and Crone, G. and Dodsworth, J. and others},
  title   = {Planck pre-launch status: The Planck mission},
  journal = {Astronomy \& Astrophysics},
  volume  = {520},
  pages   = {A1},
  year    = {2010},
  doi     = {10.1051/0004-6361/200912983}
}

@inproceedings{xgboost,
  author       = {Chen, T. and Guestrin, C.},
  title        = {XGBoost: A Scalable Tree Boosting System},
  booktitle    = {Proceedings of the 22nd ACM SIGKDD International Conference on Knowledge Discovery and Data Mining (KDD)},
  pages = {785--794},
  year         = {2016},
  doi          = {10.1145/2939672.2939785},
  publisher    = {ACM}
}

@article{Planck2016Xforeground_planck_maps,
  author = {{Planck Collaboration} and Adam, R. and Ade, P. A. R. and Aghanim, N. and Alves, M. I. R. and Armitage-Caplan, C. and Arnaud, M. and Ashdown, M. and Atrio-Barandela, F. and Aumont, J. and Baccigalupi, C. and Banday, A. J. and Barreiro, R. B.},
  title = {Planck 2015 results. X. Diffuse component separation: Foreground maps},
  journal = {Astronomy \& Astrophysics},
  volume = {594},
  pages = {A10},
  year = {2016},
  doi = {10.1051/0004-6361/201525967}
}

@article{Planck2014XII_diffuse_compo_sep,
  author = {{Planck Collaboration} and Ade, P. A. R. and Aghanim, N. and Alves, M. I. R. and Armitage-Caplan, C. and Arnaud, M. and Ashdown, M. and Atrio-Barandela, F. and Aumont, J. and Baccigalupi, C. and Banday, A. J. and Barreiro, R. B.},
  title = {Planck 2013 results. XII. Diffuse component separation},
  journal = {Astronomy \& Astrophysics},
  volume = {571},
  pages = {A12},
  year = {2014},
  doi = {10.1051/0004-6361/201321580}
}

@misc{Lundberg2018_TreeSHAP,
  author       = {Lundberg, Scott M. and Erion, Gabriel G. and Lee, Su-In},
  title        = {Consistent Individualized Feature Attribution for Tree Ensembles},
  year         = {2018},
  eprint       = {1802.03888},
  archivePrefix = {arXiv},
  primaryClass = {cs.LG},
  doi          = {10.48550/arXiv.1802.03888}
}

@inproceedings{Marzouk2025_SHAPtractability,
  author    = {Marzouk, Reda and Bassan, Shahaf and Katz, Guy and de la Higuera, Colin},
  title     = {On the Computational Tractability of the (Many) Shapley Values},
  booktitle = {Proceedings of the 28th International Conference on Artificial Intelligence and Statistics},
  volume    = {258},
  pages     = {3691--3699},
  year      = {2025},
  publisher = {PMLR}
}

@article{breiman2001random,
  author  = {Breiman, Leo},
  title   = {Random Forests},
  journal = {Machine Learning},
  year    = {2001},
  volume  = {45},
  number  = {1},
  pages   = {5--32},
  doi     = {10.1023/A:1010933404324},
  url     = {https://www.stat.berkeley.edu/~breiman/randomforest2001.pdf}
}

@article{biau2012analysis,
  author  = {Biau, G{\'e}rard},
  title   = {Analysis of a Random Forests Model},
  journal = {Journal of Machine Learning Research},
  year    = {2012},
  volume  = {13},
  pages   = {1063--1095},
  url     = {https://www.jmlr.org/papers/volume13/biau12a/biau12a.pdf}
}

@article{scornet2015consistency,
  author  = {Scornet, Erwan and Biau, G{\'e}rard and Vert, Jean-Philippe},
  title   = {Consistency of Random Forests},
  journal = {The Annals of Statistics},
  year    = {2015},
  volume  = {43},
  number  = {4},
  pages   = {1716--1741},
  doi     = {10.1214/15-AOS1321},
  url     = {https://projecteuclid.org/journals/annals-of-statistics/volume-43/issue-4/Consistency-of-random-forests/10.1214/15-AOS1321.pdf}
}

@article{Stein1987Latin_HC_sampling,
author  = {Stein, Michael},
title   = {Large Sample Properties of Simulations Using Latin Hypercube Sampling},
journal = {Technometrics},
volume  = {29},
number  = {2},
pages   = {143--151},
year    = {1987},
doi     = {10.1080/00401706.1987.10488205}
}

@article{McKay1979Latin_HC_sampling,
author  = {McKay, M. D. and Beckman, R. J. and Conover, W. J.},
title   = {Comparison of Three Methods for Selecting Values of Input Variables in the Analysis of Output from a Computer Code},
journal = {Technometrics},
volume  = {21},
number  = {2},
pages   = {239--245},
year    = {1979},
doi     = {10.1080/00401706.1979.10489755}
}

@incollection{hastie2009random,
  author    = {Hastie, Trevor and Tibshirani, Robert and Friedman, Jerome},
  title     = {Random Forests},
  booktitle = {The Elements of Statistical Learning: Data Mining, Inference, and Prediction},
  year      = {2009},
  publisher = {Springer},
  edition   = {2nd},
  chapter   = {15},
  pages     = {587--604},
  doi       = {10.1007/978-0-387-84858-7_15}
}

@article{CarrascoKind2013RF,
  author  = {Carrasco Kind, Matias and Brunner, Robert J.},
  title   = {{TPZ}: Photometric Redshift PDFs and Ancillary Information
             by Using Prediction Trees and Random Forests},
  journal = {Monthly Notices of the Royal Astronomical Society},
  volume  = {432},
  number  = {2},
  pages   = {1483--1501},
  year    = {2013},
  doi     = {10.1093/mnras/stt574}
}

@article{Draine2003_IDG,
  author    = {Draine, B. T.},
  title     = {Interstellar Dust Grains},
  journal   = {Annual Review of Astronomy and Astrophysics},
  volume    = {41},
  pages = {241--289},
  year      = {2003},
  doi       = {10.1146/annurev.astro.41.011802.094840},
  url       = {https://doi.org/10.1146/annurev.astro.41.011802.094840}
}

@article{fixsen1998lim_firas,
  author       = {Fixsen, D. J. and others},
  title        = {The Spectrum of the Extragalactic Far-Infrared Background from the COBE FIRAS Observations},
  journal      = {The Astrophysical Journal},
  volume       = {508},
  pages = {123--128},
  year         = {1998},
  doi          = {10.1086/306383},
  url          = {https://doi.org/10.1086/306383}
}

@article{planck2014overview,
  author = {{Planck Collaboration} and Ade, P. A. R. and Aghanim, N. and Alves, M. I. R. and Armitage-Caplan, C. and Arnaud, M. and Ashdown, M. and Atrio-Barandela, F. and Aumont, J. and Baccigalupi, C. and Banday, A. J. and Barreiro, R. B.},
  title = {Planck 2013 results. I. Overview of products and scientific results},
  journal = {Astronomy \& Astrophysics},
  volume = {571},
  pages = {A1},
  year = {2014},
  doi = {10.1051/0004-6361/201321529},
  url = {https://doi.org/10.1051/0004-6361/201321529}
}

@article{maffei2021bisou,
  author       = {Maffei, Bruno and Abitbol, Maxime H. and Aghanim, Nabila and Aumont, Jean and Battistelli, Ettore and Chluba, Jens and Coulon, Xavier and De Bernardis, Paolo and Douspis, Marc and Grain, Jean and Gervasoni, Stefano and Hill, John C. and Kogut, Alan and Masi, Simone and Matsumura, Takashi and O'Sullivan, Colm and Pagano, Luca and Pisano, Gianfranco and Remazeilles, Mathieu and Ritacco, Andrea and Rotti, Aditya and Sauvage, Vincent and Savini, Giorgio and Stever, Samuel L. and Tartari, Andrea and Thiele, Lars and Trappe, Nicolas},
  title        = {BISOU: a balloon project to measure the CMB spectral distortions},
  journal      = {arXiv:2111.00246},
  year         = {2021},
  url          = {https://arxiv.org/abs/2111.00246}
}

@article{chluba2021new_horizons,
  author       = {Chluba, Jens and Abitbol, Maximilian H. and Aghanim, Nabila and Ali-Haïmoud, Yacine and Alvarez, Marcelo and Basu, Kaustuv and Bolliet, Boris and Burigana, Chiara and de Bernardis, Paolo and Delabrouille, Jacques and Dimastrogiovanni, Emanuela and Finelli, Fabio and Fixsen, Dale J. and Hart, Luke and Hernández-Monteagudo, Carlos and Hill, J. Colin and Kogut, Alan and Kohri, Kazunori and Lesgourgues, Julien and Maffei, Bruno and Mather, John C. and Mukherjee, Suvodip and Patil, Suvodip P. and Ravenni, Andrea and Remazeilles, Mathieu and Rotti, Aditya and Rubiño-Martín, Jos\'{e} Alberto and Silk, Joseph and Sunyaev, Rashid A. and Switzer, Eric R.},
  title        = {New horizons in cosmology with spectral distortions of the cosmic microwave background},
  journal      = {Experimental Astronomy},
  volume       = {51},
  number = {3},
  pages = {1515--1554},
  year         = {2021},
  doi          = {10.1007/s10686-021-09729-5},
  eprint       = {arXiv:1909.01593},
  archivePrefix = {arXiv},
  primaryClass = {astro-ph.CO},
  url          = {https://doi.org/10.1007/s10686-021-09729-5}
}

@article{kogut2012pixie,
  author       = {Kogut, Alan and Fixsen, Dale J. and Chuss, David T. and Dotson, Jessie and Dwek, Eli and Halpern, Mark and Hinshaw, Gary F. and Meyer, Steve M. and Moseley, Samuel H. and Seiffert, Michael D. and Spergel, David N. and Wollack, Edward J.},
  title        = {The Primordial Inflation Explorer (PIXIE): a nulling polarimeter for cosmic microwave background observations},
  journal      = {Experimental Astronomy},
  volume       = {36},
  number = {1--2},
  pages = {145--183},
  year         = {2013},
  month        = {Aug},
  doi          = {10.1007/s10686-013-9335-7},
  url          = {https://doi.org/10.1007/s10686-013-9335-7},
}

@article{kogut2016pixie,
  author       = {Kogut, Alan and Chluba, Jens and Fixsen, Dale J. and Meyer, Steve M. and Spergel, David N. and Dunkley, Joanna and Bennett, Charles L. and Crites, Abigail T. and Erickson, Nathan and Essinger-Hileman, Thomas and Fialkov, Anastasia and Hill, J. Colin and Hinshaw, Gary F. and Johnson, Bradley R. and Lawrence, Charles R. and Mather, John C. and Moseley, Samuel H. and Paoletti, Daniela and Partridge, R. Bruce and Rotti, Aditya and Seiffert, Michael D. and Stevenson, Mark and Switzer, Eric R. and Wollack, Edward J.},
  title        = {The Primordial Inflation Explorer (PIXIE): a nulling polarimeter for cosmic microwave background observations},
  journal      = {Journal of Cosmology and Astroparticle Physics},
  volume       = {2016},
  number = {08},
  pages = {025},
  year         = {2016},
  doi          = {10.1088/1475-7516/2016/08/025},
  url          = {https://doi.org/10.1088/1475-7516/2016/08/025},
}

@misc{pristine_ias,
  author = {Aghanim, Nabila and Abitbol, Maximilian and Aumont, Jonathan and Battistelli, Elia and Benabed, Karim and Bouchet, François and Boulanger, François and Casoli, Fabienne and Charles, Ivan and Chluba, Jens and Cuttaia, Francesco and Dole, Hervé and Douspis, Marian and Durrer, Ruth and Duval, Jean-Marc and Eriksen, Hans Kristian and Finelli, Fabio and Fixsen, Dale and Galli, Silvia and Gervasi, Massimo and Grain, Julien and Guillet, Vincent and Haziot, Ariel and Hernandez-Monteagudo, Carlos and Kogut, Al and Kunz, Martin and Lagache, Guilaine and Langer, Mathieu and Macias-Perez, Juan and Maffei, Bruno and Maillard, Jean-Pierre and Mangilli, Anna and Masi, Silvia and Nati, Federico and Montaruli, Teresa and O'Sullivan, Creidhe and Pagano, Luca and Paoletti, Daniela and Philippon, Anne and Pisano, Giampaolo and Pitrou, Cyril and Pointecouteau, Etienne and Puget, Jean-Loup and Remazeilles, Mathieu and Rodriguez, Louis and Roussafi, Abdellah and Rubino-Martin, Jose-Alberto and Savini, Giorgio and Silk, Joseph and Starck, Jean-Luc and Sunyaev, Rashid and Tartari, Andrea and Terenzi, Luca and Trappe, Neil and Triqueneaux, Sébastien and Tucker, Carole and Vermeulen, Gérard and Wehus, Ingunn and Winter, Berend},
  title  = {PRISTINE: Polarized Radiation Interferometer for Spectral disTortions and Inflation Exploration},
  note   = {Mission proposée dans l'appel F de l’ESA, programme Cosmic Vision 2015--2025, Institut d'Astrophysique Spatiale, Université Paris-Saclay},
  url    = {https://www.ias.u-psud.fr/en/content/pristine},
  year   = {2018}
}

@article{Thorne2017pysm,
  author = {Thorne, Benjamin and Dunkley, Jo and Alonso, David and Naess, Sigurd},
  title = {The Python Sky Model: software for simulating the Galactic microwave sky},
  journal = {Journal of Cosmology and Astroparticle Physics},
  volume = {2017},
  number = {04},
  pages = {014},
  year = {2017},
  doi = {10.1088/1475-7516/2017/04/014}
}

@article{Zonca2021pysm3,
  author = {Zonca, Andrea and Singer, L. Philip and Lenz, Daniel and Reinecke, Martin and Rosset, Cl\'{e}ment and Hivon, Eric and Górski, Krzysztof M.},
  title = {The Python Sky Model 3 software},
  journal = {Journal of Open Source Software},
  volume = {6},
  number = {67},
  pages = {3783},
  year = {2021},
  doi = {10.21105/joss.03783}
}

@ARTICLE{Haslam2001_Galactic_synchrotron_foreground,
       author = {{Haslam}, C.~G.~T. and others},
        title = "{Galactic Synchrotron Radiation}",
      journal = {Annual Review of Astronomy and Astrophysics},
       volume = {39},
        pages = {249--290},
         year = 2001,
          doi = {10.1146/annurev.astro.39.1.249}
}

@ARTICLE{Planck2020_CMB_foreground_components,
       author = {{Planck Collaboration}},
        title = "{Planck 2018 results. I. Overview and the cosmological legacy of Planck}",
      journal = {\aap},
       volume = {641},
        pages = {A1},
         year = 2020,
          doi = {10.1051/0004-6361/201833880},
archivePrefix = {arXiv},
       eprint = {1807.06205}
}

@article{Dickinson2003ff_template,
  author = {Dickinson, C. and Davies, R. D. and Davis, R. J.},
  title = {Towards a free-free template for CMB foregrounds},
  journal = {Monthly Notices of the Royal Astronomical Society},
  volume = {341},
  pages = {369--384},
  year = {2003},
  doi = {10.1046/j.1365-8711.2003.06456.x}
}

@book{Draine2011FreeFree,
  author = {Draine, B. T.},
  title = {Physics of the Interstellar and Intergalactic Medium},
  publisher = {Princeton University Press},
  address = {Princeton, NJ},
  year = {2011},
  isbn = {978-0-691-12214-4},
  note = {See chapter on free-free Galactic emission}
}

@ARTICLE{Oliveira1998,
       author = {{de Oliveira-Costa}, A. and {Tegmark}, Max and {Page}, Lyman A. and {Boughn}, Stephen P.},
        title = "{Galactic Emission at 19 GHZ}",
      journal = {\apjl},
         year = 1998,
        month = dec,
       volume = {509},
       number = {1},
        pages = {L9-L12},
          doi = {10.1086/311754},
archivePrefix = {arXiv},
       eprint = {astro-ph/9807329},
 primaryClass = {astro-ph},
       adsurl = {https://ui.adsabs.harvard.edu/abs/1998ApJ...509L...9D}
}

@ARTICLE{Battistelli2019,
       author = {{Battistelli}, E.~S. and {Fatigoni}, S. and {Murgia}, M. and {Buzzelli}, A. and {Carretti}, E. and {Castangia}, P. and {Concu}, R. and {Cruciani}, A. and {de Bernardis}, P. and {Genova-Santos}, R. and {Govoni}, F. and {Guidi}, F. and {Lamagna}, L. and {Luzzi}, G. and {Masi}, S. and {Melis}, A. and {Paladini}, R. and {Piacentini}, F. and {Poppi}, S. and {Radiconi}, F. and {Rebolo}, R. and {Rubino-Martin}, J.~A. and {Tarchi}, A. and {Vacca}, V.},
        title = "{Strong Evidence of Anomalous Microwave Emission from the Flux Density Spectrum of M31}",
      journal = {\apjl},
         year = 2019,
        month = jun,
       volume = {877},
       number = {2},
          eid = {L31},
        pages = {L31},
          doi = {10.3847/2041-8213/ab21de},
archivePrefix = {arXiv},
       eprint = {1905.12276},
 primaryClass = {astro-ph.GA},
       adsurl = {https://ui.adsabs.harvard.edu/abs/2019ApJ...877L..31B}
}

@ARTICLE{Lagache2003,
       author = {{Lagache}, G.},
        title = "{The large-scale anomalous microwave emission revisited by WMAP.}",
      journal = {\aap},
         year = 2003,
        month = jul,
       volume = {405},
        pages = {813-819},
          doi = {10.1051/0004-6361:20030545},
archivePrefix = {arXiv},
       eprint = {astro-ph/0303335},
 primaryClass = {astro-ph},
       adsurl = {https://ui.adsabs.harvard.edu/abs/2003A&A...405..813L}
}

@ARTICLE{Arce2020,
       author = {{Arce-Tord}, C. and {Vidal}, Matias and {Casassus}, Simon and {C{\'a}rcamo}, Miguel and {Dickinson}, Clive and {Hensley}, Brandon S. and {G{\'e}nova-Santos}, Ricardo and {Bond}, J. Richard and {Jones}, Michael E. and {Readhead}, Anthony C.~S. and {Taylor}, Angela C. and {Zensus}, J. Anton},
        title = "{Resolved observations at 31 GHz of spinning dust emissivity variations in {\ensuremath{\rho}} Oph}",
      journal = {\mnras},
         year = 2020,
        month = jan,
       volume = {495},
       number = {3},
        pages = {3482-3493},
          doi = {10.1093/mnras/staa1422},
archivePrefix = {arXiv},
       eprint = {1910.06359},
 primaryClass = {astro-ph.GA},
       adsurl = {https://ui.adsabs.harvard.edu/abs/2020MNRAS.495.3482A}
}

@article{Kogut1996ame_fir,
  author = {Kogut, A. and Banday, A. J. and Bennett, C. L. and others},
  title = {High-Latitude Galactic Emission in the COBE Differential Microwave Radiometers First-Year Sky Maps},
  journal = {The Astrophysical Journal},
  volume = {464},
  pages = {L5--L9},
  year = {1996},
  doi = {10.1086/310067}
}

@article{Planck2011XXame_new,
  author = {{Planck Collaboration} and Ade, P. A. R. and Aghanim, N. and Armitage-Caplan, C. and Arnaud, M. and Ashdown, M. and Atrio-Barandela, F. and Aumont, J. and Baccigalupi, C. and Banday, A. J. and Barreiro, R. B.},
  title = {Planck early results. XX. New light on anomalous microwave emission from spinning dust grains},
  journal = {Astronomy \& Astrophysics},
  volume = {536},
  pages = {A20},
  year = {2011},
  doi = {10.1051/0004-6361/201116470}
}

@article{Lagache2003cib,
  author = {Lagache, G. and Dole, H. and Puget, J.-L.},
  title = {Modelling the infrared galaxy evolution using the Cosmic Infrared Background},
  journal = {Monthly Notices of the Royal Astronomical Society},
  volume = {338},
  pages = {555--571},
  year = {2003},
  doi = {10.1046/j.1365-8711.2003.06013.x}
}

@article{Dole2004cib,
  author = {Dole, H. and Lagache, G. and Puget, J.-L. and others},
  title = {The Cosmic Infrared Background Resolved by Spitzer. Contributions of Mid-Infrared Galaxies to the Far-Infrared Background},
  journal = {The Astrophysical Journal Supplement Series},
  volume = {154},
  pages = {93--97},
  year = {2004},
  doi = {10.1086/422717}
}

@article{Puget1996cib_COBE,
  author = {Puget, J.-L. and Abergel, A. and Bernard, J.-P. and Boulanger, F. and Burton, W. B. and Desert, F.-X. and Hartmann, D.},
  title = {Tentative detection of a cosmic far-infrared background with COBE},
  journal = {Astronomy \& Astrophysics},
  volume = {308},
  pages = {L5--L8},
  year = {1996},
  doi = {10.48550/arXiv.astro-ph/9511042}
}

@article{Masi2021_COSMO,
  author = {Masi, S. and Battistelli, E. and de Bernardis, P. and Coppolecchia, A. and Columbro, F. and D'Alessandro, G. and De Petris, M. and Lamagna, L. and Marchitelli, E. and Mele, L. and Paiella, A. and Piacentini, F. and Pisano, G. and Bersanelli, M. and Franceschet, C. and Manzan, E. and Mennella, D. and Realini, S. and Cibella, S. and Martini, F. and Pettinari, G. and Coppi, G. and Gervasi, M. and Limonta, A. and Zannoni, M. and Piccirillo, L. and Tucker, C.},
  title = {The COSmic Monopole Observer (COSMO)},
  journal = {arXiv e-prints},
  year = {2021},
  volume = {arXiv:2110.12254},
  url = {https://arxiv.org/abs/2110.12254}
}

@article{Bennett2003_sync_wmap,
  author = {Bennett, C. L. and Hill, R. S. and Hinshaw, G. and Nolta, M. R. and Odegard, N. and Page, L. and Spergel, D. N. and Weiland, J. L. and Jarosik, N. and Kogut, A. and Limon, M. and Meyer, S. S. and Tucker, G. S. and Wollack, E.},
  title = {First Year Wilkinson Microwave Anisotropy Probe (WMAP) Observations: Foreground Emission},
  journal = {Astrophysical Journal Supplement Series},
  volume = {148},
  pages = {97--117},
  year = {2003},
  doi = {10.1086/377253},
  tag = {sync_WMAP}
}

@article{Friedman2001_CML_boosting,
  author = {Friedman, J. H.},
  title = {Greedy function approximation: A gradient boosting machine},
  journal = {The Annals of Statistics},
  volume = {29},
  number = {5},
  pages = {1189--1232},
  year = {2001},
  doi = {10.1214/aos/1013203451}
}

@inproceedings{Caruana2006_SupervisedLearningComparison,
  author = {Caruana, Rich and Niculescu-Mizil, Alexandru},
  title = {An Empirical Comparison of Supervised Learning Algorithms},
  booktitle = {Proceedings of the 23rd International Conference on Machine Learning (ICML 2006)},
  year = {2006},
  doi = {10.1145/1143844.1143865}
}

@article{Strobl2008_CondVarImpRF,
  author = {Strobl, Carolin and Boulesteix, Anne-Laure and Kneib, Thomas and Augustin, Thomas and Zeileis, Achim},
  title = {Conditional variable importance for random forests},
  journal = {BMC Bioinformatics},
  volume = {9},
  pages = {307},
  year = {2008},
  doi = {10.1186/1471-2105-9-307}
}

@article{LiteBIRD2023_PTEP,
  author = {{LiteBIRD Collaboration} and Allys, E. and Arnold, K. and Aumont, J. and Aurlien, R. and Azzoni, S. and Baccigalupi, C. and Banday, A. J. and Banerji, R. and Barreiro, R. B. and Bartolo, N. and Bautista, L. and Beck, D. and Beckman, S. and Bersanelli, M. and Boulanger, F. and Brilenkov, M. and Bucher, M. and Calabrese, E. and Campeti, P. and Carones, A. and Casas, F. J. and Catalano, A. and Chan, V. and Cheung, K. and Chinone, Y. and Clark, S. E.},
  title = {Probing cosmic inflation with the LiteBIRD cosmic microwave background polarization survey},
  journal = {Progress of Theoretical and Experimental Physics},
  volume = {2023},
  number = {4},
  pages = {042F01},
  year = {2023},
  doi = {10.1093/ptep/ptac150}
}

@article{Mather1990_COBE_FIRAS,
  author = {Mather, J. C. and Cheng, E. S. and Cottingham, D. A. and Eplee, R. E. and Fixsen, D. J. and Hewagama, T. and Isaacman, R. and Jensen, K. A. and Meyer, S. S. and Noerdlinger, P. D. and Shafer, R. A. and Smoot, G. F. and Weiss, R.},
  title = {A Preliminary Measurement of the Cosmic Microwave Background Spectrum by the Cosmic Background Explorer (COBE) Satellite},
  journal = {Astrophysical Journal},
  volume = {354},
  pages = {L37--L40},
  year = {1990},
  doi = {10.1086/185714}
}

@article{Martin1970_martin_puplett,
  author = {Martin, D. H. and Puplett, E.},
  title = {Polarised interferometric spectrometry for the millimetre and submillimetre wavebands},
  journal = {Infrared Physics},
  volume = {10},
  pages = {105--109},
  year = {1970},
  doi = {10.1016/0020-0891(70)90092-5}
}

@article{Planck2014zodi,
  author = {{Planck Collaboration} and Ade, P. A. R. and Aghanim, N. and Alves, M. I. R. and Armitage-Caplan, C. and Arnaud, M. and Ashdown, M. and Atrio-Barandela, F. and Aumont, J. and Baccigalupi, C. and Banday, A. J. and Barreiro, R. B.},
  title = {Planck 2013 results. XIV. Zodiacal emission},
  journal = {Astronomy \& Astrophysics},
  volume = {571},
  pages = {A14},
  year = {2014},
  doi = {10.1051/0004-6361/201321543},
  eprint = {1303.5074}
}

@article{Kogut2024PIXIE,
  author = {Kogut, Alan and Switzer, Eric and Fixsen, Dale and Aghanim, Nabila and Chluba, Jens and Chuss, David T. and Delabrouille, Jacques and Dvorkin, Cora and Hensley, Brandon and Hill, J. Colin and Maffei, Bruno and Pullen, Anthony R. and Rotti, Aditya and Sabyr, Alina and Thiele, Leander and Wollack, Edward J. and Zelko, Ioana},
  title = {The Primordial Inflation Explorer (PIXIE): Mission Design and Science Goals},
  journal = {arXiv e-prints},
  year = {2024},
  pages = {arXiv:2405.20403},
  url = {https://arxiv.org/abs/2405.20403}
}

@article{Murphy2010_multimode_planck,
  author  = {Murphy, J. A. and Peacocke, T. and Maffei, B. and McAuley, I. and Noviello, F. and Yurchenko, V. and Ade, P. A. R. and Savini, G. and Lamarre, J.-M. and Brossard, J. and Colgan, R. and Gleeson, E. and Lange, A. E. and Longval, Y. and Pisano, G. and Puget, J.-L. and Ristorcelli, I. and Sudiwala, R. and Wylde, R. J.},
  title   = {Multi‐mode horn design and beam characteristics for the Planck satellite},
  journal = {Journal of Instrumentation},
  volume  = {5},
  pages = {T04001:1--21},
  year    = {2010},
  doi     = {10.1088/1748-0221/5/04/T04001}
}

@article{Mashian2016_CMB_CO,
  author    = {Mashian, Natalie and Loeb, Abraham and Sternberg, Amiel},
  title     = {Spectral distortion of the CMB by the cumulative CO emission from galaxies throughout cosmic history},
  journal   = {Monthly Notices of the Royal Astronomical Society Letters},
  volume    = {458},
  number    = {1},
  pages     = {L99--L103},
  year      = {2016},
  doi       = {10.1093/mnrasl/slw086},
  url       = {https://doi.org/10.1093/mnrasl/slw086}
}

@article{Krick2012_ZodiacalIRAC,
  author    = {Krick, Jessica E. and Glaccum, William J. and Carey, Sean J. and Lowrance, Patrick J. and Surace, Jason A. and Ingalls, James G. and Hora, Joseph L. and Reach, William T.},
  title     = {A Spitzer/IRAC Measure of the Zodiacal Light},
  journal   = {The Astrophysical Journal},
  volume    = {754},
  number    = {1},
  pages     = {53},
  year      = {2012},
  doi       = {10.1088/0004-637X/754/1/53},
  url       = {https://doi.org/10.1088/0004-637X/754/1/53}
}

@inproceedings{Maffei2024_BISOU_pathfinder,
  author = {Maffei, B. and Aghanim, N. and Aumont, J. and Battistelli, E. and Beelen, A. and Besnard, A. and Borgo, B. and Calvo, M. and Catalano, A. and Chluba, J. and Coulon, X. and De Bernardis, P. and de Jabrun, C. and Douspis, M. and Errard, J. and Grain, J. and Guiot, P. and Hill, J. C. and Ishino, H. and Kogut, A. and Lagache, G. and Macias-Perez, J. and Masi, S. and Matsumura, T. and Monfardini, A. and O'Sullivan, C. and Pagano, L. and Patanchon, G. and Pisano, G. and Pitre, L. and Ponthieu, N. and Remazeilles, M. and Ritacco, A. and Savini, G. and Sauvage, V. and Shitov, A. and Stever, S. L. and Tartari, A. and Thiele, L. and Trappe, N. and Aubrun, J.-F. and Bray, N. and Louvel, S.},
  title = {BISOU: A balloon pathfinder for CMB spectral distortions studies},
  booktitle = {Millimeter, Submillimeter, and Far-Infrared Detectors and Instrumentation for Astronomy XII},
  year = {2024},
  doi = {10.1117/12.3018371},
  url = {https://doi.org/10.1117/12.3018371}
}

@article{Werner2004_Spitzer,
  author    = {Werner, M. W. and Roellig, T. L. and Low, F. J. and Rieke, G. H. and Rieke, M. J. and Smith, H. A. and Chary, R. R. and Gehrz, R. D. and Soifer, B. T. and Wright, E. L.},
  title     = {The Spitzer Space Telescope Mission},
  journal   = {The Astrophysical Journal Supplement Series},
  volume    = {154},
  pages     = {1--9},
  year      = {2004},
  doi       = {10.1086/423746},
  url       = {https://doi.org/10.1086/423746}
}

@article{Wright1991co_mw,
  author = {Wright, E. L. and Mather, J. C. and Fixsen, D. J. and others},
  title = {Interpretation of the Cosmic Far-Infrared Background from COBE Observations},
  journal = {The Astrophysical Journal},
  volume = {381},
  pages = {200--208},
  year = {1991},
  doi = {10.1086/170587}
}

@article{fixsen2009cmbSD,
  author       = {Fixsen, D. J.},
  title        = {The Temperature of the Cosmic Microwave Background},
  journal      = {The Astrophysical Journal},
  volume       = {707},
  number = {2},
  pages = {916--920},
  year         = {2009},
  doi          = {10.1088/0004-637X/707/2/916},
  url          = {https://doi.org/10.1088/0004-637X/707/2/916}
}

@misc{aghanim2022fossil,
  author       = {Aghanim, Nabila and Abitbol, Maxime H. and Aumont, Jonathan and Battistelli, Elia and Bolliet, Boris and Chluba, Jens and Coulon, Xavier and de Bernardis, Paolo and Douspis, Marc and Finelli, Fabio and Fixsen, Dale and Hernández-Monteagudo, Carlos and Hill, J. Colin and Kogut, Alan and Kuruvilla, Joseph and Lagache, Guilaine and Macías-P\'{e}rez, Juan and Maffei, Bruno and Maillard, Jean-Pierre and Masi, Silvia and Monfardini, Alessandro and O'Sullivan, Creidhe and Pagano, Luca and Pitrou, Cyril and Ponthieu, Nicolas and Rotti, Aditya and Rubiño-Martín, Jos\'{e} Alberto and Savini, Giorgio and Switzer, Eric and Tanimura, Hideki and Tartari, Andrea and Thiele, Leander and Trappe, Neil and ...},
  title        = {FOSSIL: FTS for CMB Spectral Distortion Exploration --- A mission concept for the ESA M-class call},
  year         = {2022},
  url          = {https://www.ias.u-psud.fr/sites/default/files/FOSSIL-web_0.pdf},
}

@article{chluba2017moment_expansion,
  author       = {Chluba, Jens and Hill, J. Colin and Abitbol, Mark H.},
  title        = {Moment expansion for the analysis of CMB spectral distortions},
  journal      = {Monthly Notices of the Royal Astronomical Society},
  volume       = {472},
  number = {2},
  pages = {1195--1207},
  year         = {2017},
  doi          = {10.1093/mnras/stx2018},
  url          = {https://doi.org/10.1093/mnras/stx2018}
}

@article{ysard2024themis2,
  title = {THEMIS 2.0: A self-consistent model for dust extinction, emission, and polarisation},
  author = {Ysard, N. and Jones, A.~P. and Guillet, V. and Demyk, K. and Decleir, M. and Verstraete, L. and Choubani, I. and Miville-Desch{\^e}nes, M.-A. and Fanciullo, L.},
  journal = {Astronomy \& Astrophysics},
  volume = {684},
  pages = {A34},
  year = {2024},
  doi = {10.1051/0004-6361/202348391}
}

@article{hensley2022astrodust,
  title = {{The Astrodust+PAH Model: A Unified Description of the Extinction, Emission, and Polarization from Dust in the Diffuse Interstellar Medium}},
  author = {Hensley, Brandon S. and Draine, B. T.},
  journal = {Astrophysical Journal},
  year = {2022},
  note = {arXiv:2208.12365}
}

@inproceedings{shap,
  author       = {Lundberg, Scott M. and Lee, Su-In},
  title        = {A Unified Approach to Interpreting Model Predictions},
  booktitle    = {Advances in Neural Information Processing Systems (NIPS) 30},
  pages = {4765--4774},
  year         = {2017},
  url          = {https://arxiv.org/abs/1705.07874},
  note         = {arXiv preprint arXiv:1705.07874}
}

@incollection{Shapley1953_ShapleyValue,
  author    = {Shapley, Lloyd S.},
  title     = {A Value for $n$-Person Games},
  booktitle = {Contributions to the Theory of Games II},
  editor    = {Kuhn, Harold W. and Tucker, Albert W.},
  pages     = {307--317},
  year      = {1953},
  publisher = {Princeton University Press},
  address   = {Princeton}
}

@incollection{Winter2002_ShapleyValue,
  author    = {Winter, Eyal},
  title     = {The Shapley Value},
  booktitle = {Handbook of Game Theory with Economic Applications},
  volume    = {3},
  pages     = {2025--2054},
  year      = {2002},
  publisher = {Elsevier},
  doi       = {10.1016/S1574-0005(02)03016-3}
}

@article{sklearn_pedregosa2011,
  title={Scikit-learn: Machine learning in Python},
  author={Pedregosa, Fabian and Varoquaux, Ga{\"e}l and Gramfort, Alexandre and Michel, Vincent and Thirion, Bertrand and Grisel, Olivier and Blondel, Mathieu and Prettenhofer, Peter and Weiss, Ron and Dubourg, Vincent and others},
  journal={Journal of machine learning research},
  volume={12},
  number = {Oct},
  pages = {2825--2830},
  year={2011}
}

\begin{appendix}




\section{Validation of the \texttt{XGBoost} model}
\label{app:robustness}

\begin{figure}[!ht]
    \centering
    \includegraphics[height=6cm]{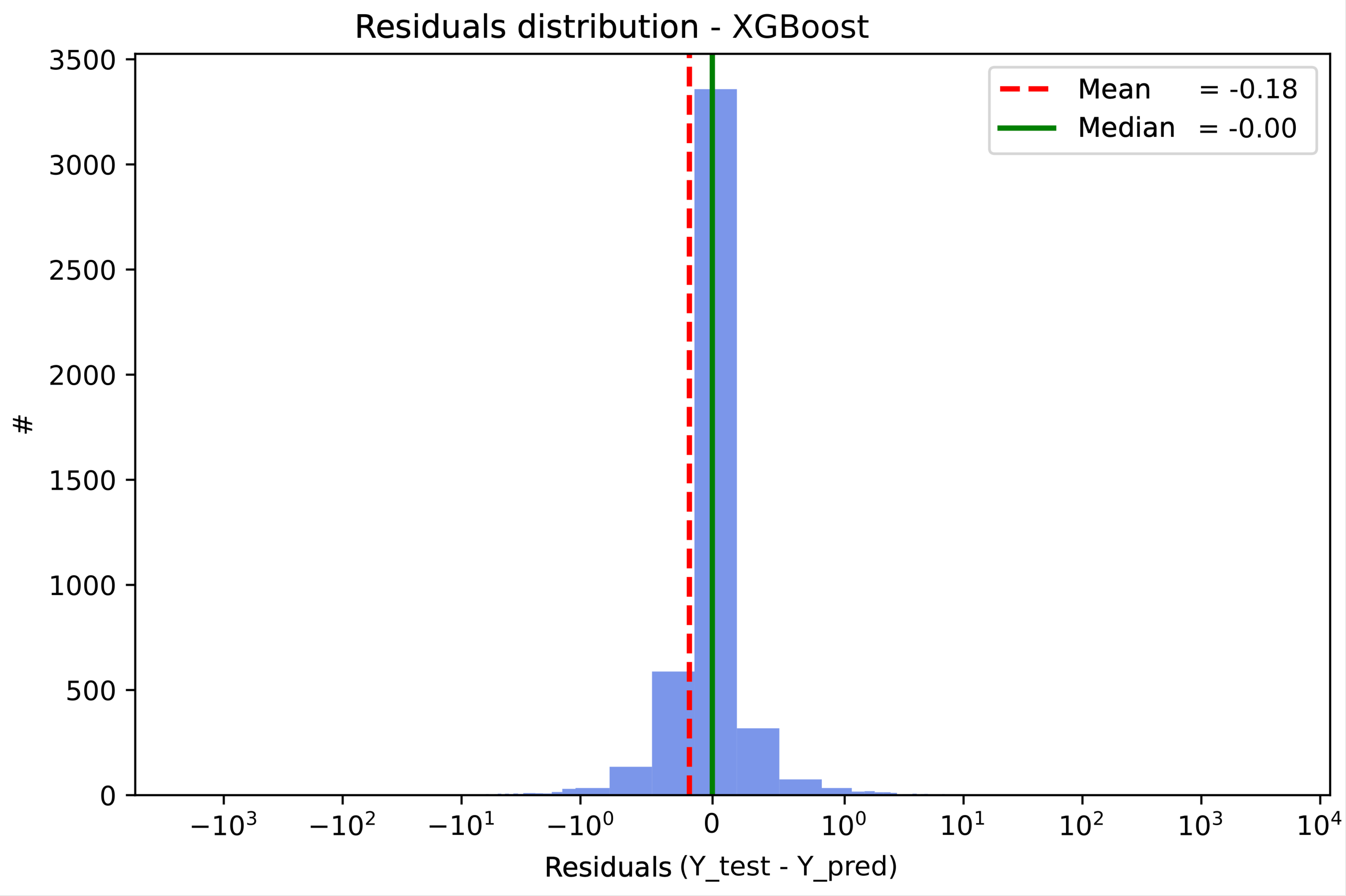}
    \caption{Residual distribution on the test set for the $y$-SNR prediction.}
    \label{fig: hist SHAP XGB}
\end{figure}

The robustness of the \texttt{XGBoost}-based analysis is assessed using
complementary diagnostics of the model predictions and training behaviour.
These include the residual distribution, the Mean Absolute Error (MAE),
the Root Mean Square Error (RMSE), the learning curves, and the
consistency between the SHAP decomposition and the model predictions.
Together, these diagnostics provide complementary information on the
predictive accuracy, stability, and interpretability of the model, and
support the SHAP analysis presented in Sect.~\ref{subsec: robustness_global}.
For the $y$-SNR prediction, the model achieves $R^2=0.927$, with a mean
residual of $-0.178$, a MAE of $0.314$, and a RMSE of $2.732$
(see Table~\ref{tab : err XGBoost y}). Figure~\ref{fig: hist SHAP XGB} shows that the residual distribution is centred close to zero, indicating no significant global bias. The MAE
corresponds to approximately $1.5\%$ of a typical $y$-SNR of $\sim20$,
while the larger RMSE reflects a small number of relatively large
deviations. These outliers represent approximately $1\%$ of the test
sample and occur predominantly at very high predicted $y$-SNR values
($\gtrsim1000\,\sigma$), indicating that the approximation is less
accurate in this limited region of the parameter space.

\begin{table}[h!]
\centering
\caption{Performance metrics of the \texttt{XGBoost} model for the $y$ observable.}
\begin{tabular}{l c}
\hline
Metric & Value \\
\hline
$R^2$ & 0.927 \\
Mean residual & $-0.178$ \\
MAE & $0.314$ \\
RMSE & $2.732$ \\
\hline
\end{tabular}
\label{tab : err XGBoost y}
\end{table}

\begin{figure}[!ht]
    \centering
    \includegraphics[height=6cm]{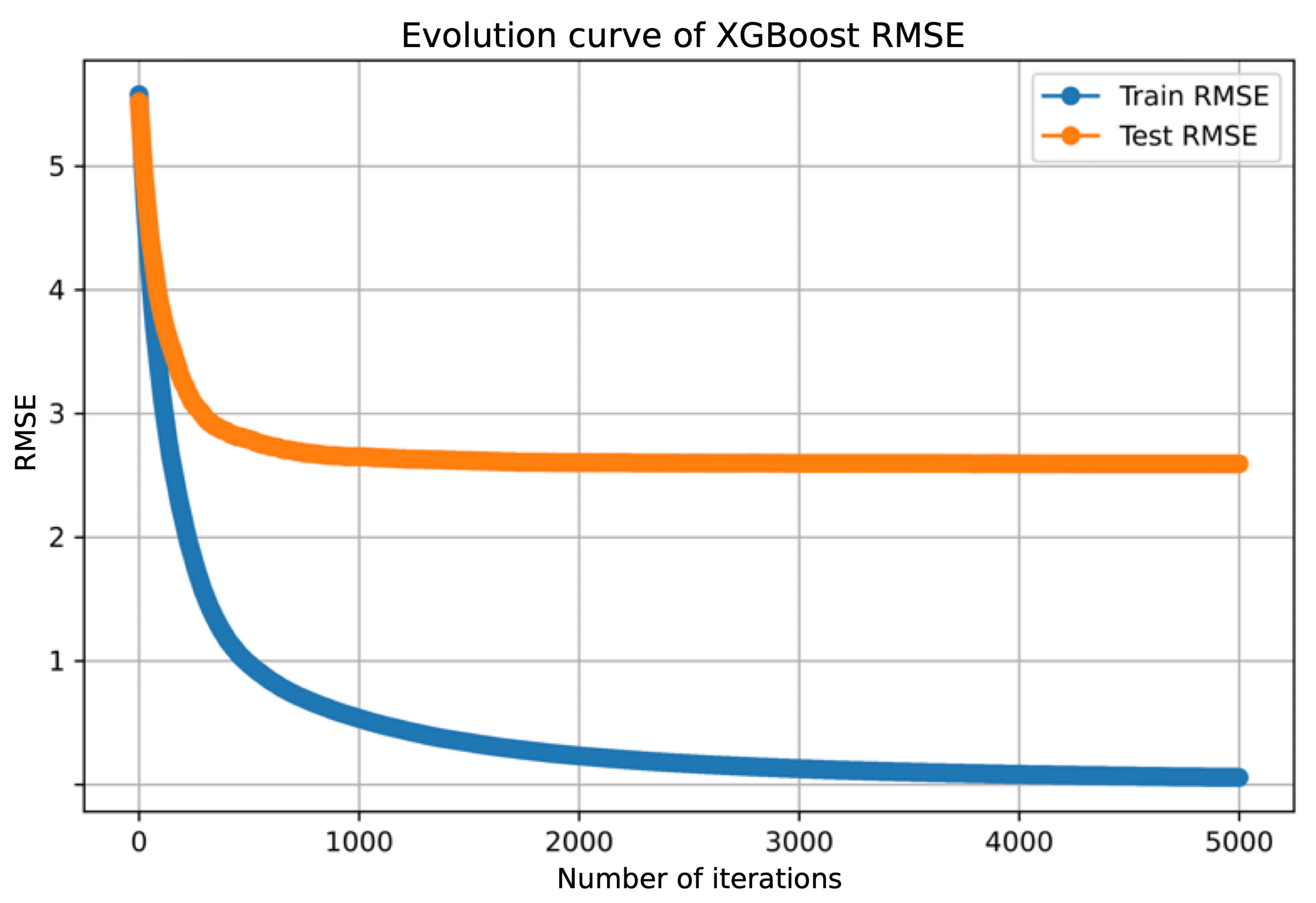}
    \caption{Learning curves showing the RMSE evolution for training and
    test sets as a function of boosting iterations.}
    \label{fig: err SHAP XGB}
\end{figure}

The learning curves shown in Fig.~\ref{fig: err SHAP XGB} provide an
additional diagnostic of the model behaviour during training. The RMSE
decreases with the number of boosting iterations and reaches a stable
regime for both the training and test samples. The absence of a
significant divergence between the two curves provides no clear evidence
of overfitting over the range considered. Together with the residual
analysis, these results support the use of the trained model to explore
the dependence of the predicted sensitivity on the instrumental
parameters.

\begin{figure}[!ht]
    \centering
    \includegraphics[height=7cm]{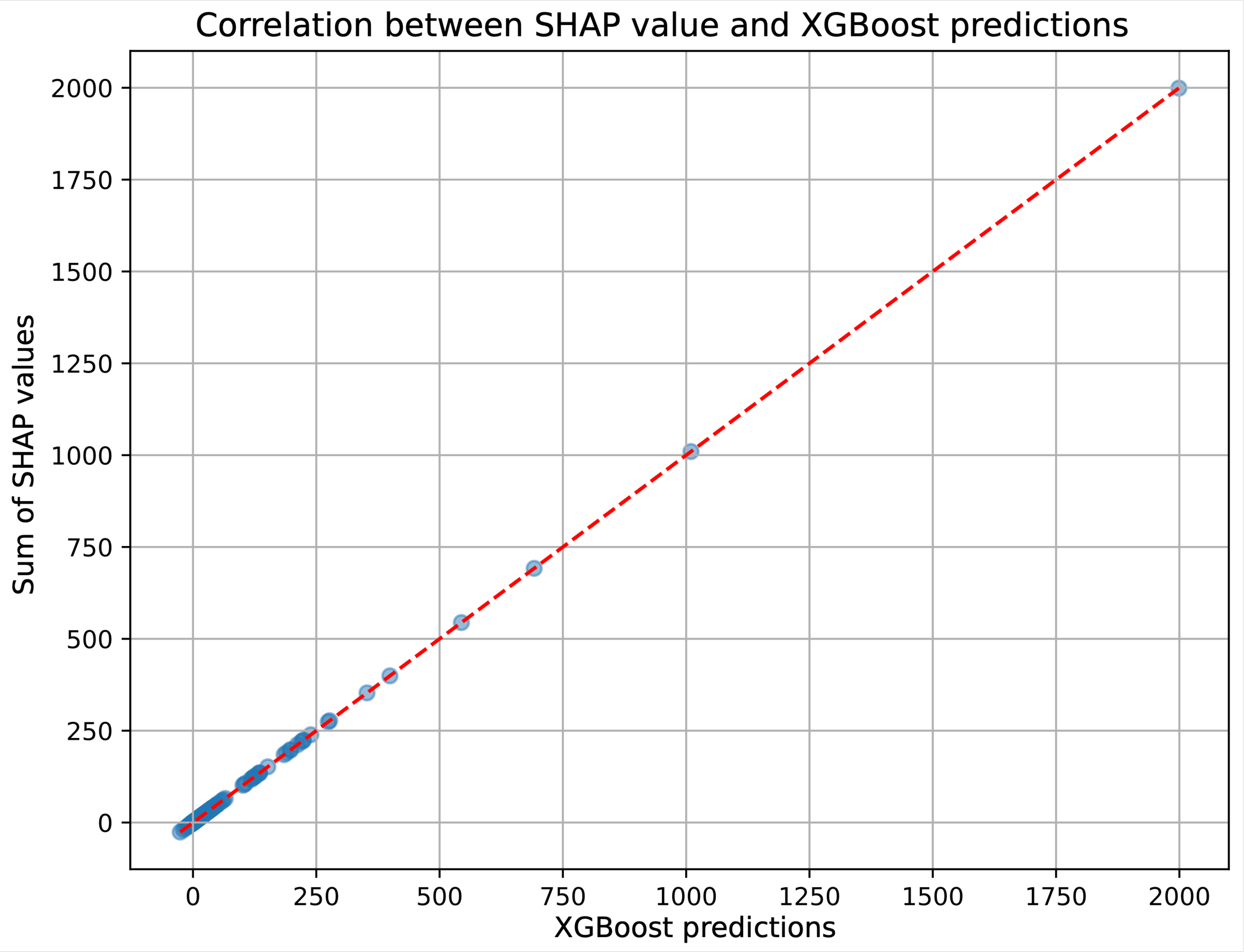}
    \caption{Consistency check between the SHAP decomposition and
    \texttt{XGBoost} predictions. The summed SHAP contributions reproduce
    the model outputs; the red line indicates the expected one-to-one
    relation.}
    \label{fig: corr SHAP XGB}
\end{figure}

The SHAP decomposition is then compared directly with the
\texttt{XGBoost} predictions on the test set. As shown in
Fig.~\ref{fig: corr SHAP XGB}, the sum of the SHAP contributions
reproduces the model predictions with the expected one-to-one relation.
This confirms the internal consistency of the SHAP decomposition and
shows that the feature contributions correctly account for the output of
the trained model. This validation is particularly important because the SHAP values are subsequently used to interpret the dependence of the predicted sensitivity on the instrumental parameters.

The isolated positive SHAP contribution associated with low $\eta$ is
therefore interpreted as a local limitation of the learned approximation,
rather than as a failure of the SHAP method. SHAP values explain the
predictions of the underlying machine-learning model and do not impose
the physical properties of the instrument. A locally inaccurate
approximation of the physical mapping can consequently produce a
locally non-physical feature attribution even when the global predictive
performance remains high. This illustrates the importance of combining
SHAP-based interpretation with predictive diagnostics and physical
consistency checks when applying machine-learning models to instrument
optimisation.

\onecolumn
\section{Additional SHAP analyses}
\label{app:shap}

\begin{center}

\includegraphics[height=6cm]{figures/SHAP_MU.pdf}

\captionof{figure}{Same as Fig.~\ref{fig: SHAP y}, but for $\mu$.}
\label{fig: SHAP mu}

\vspace{0.5cm}

\includegraphics[height=6cm]{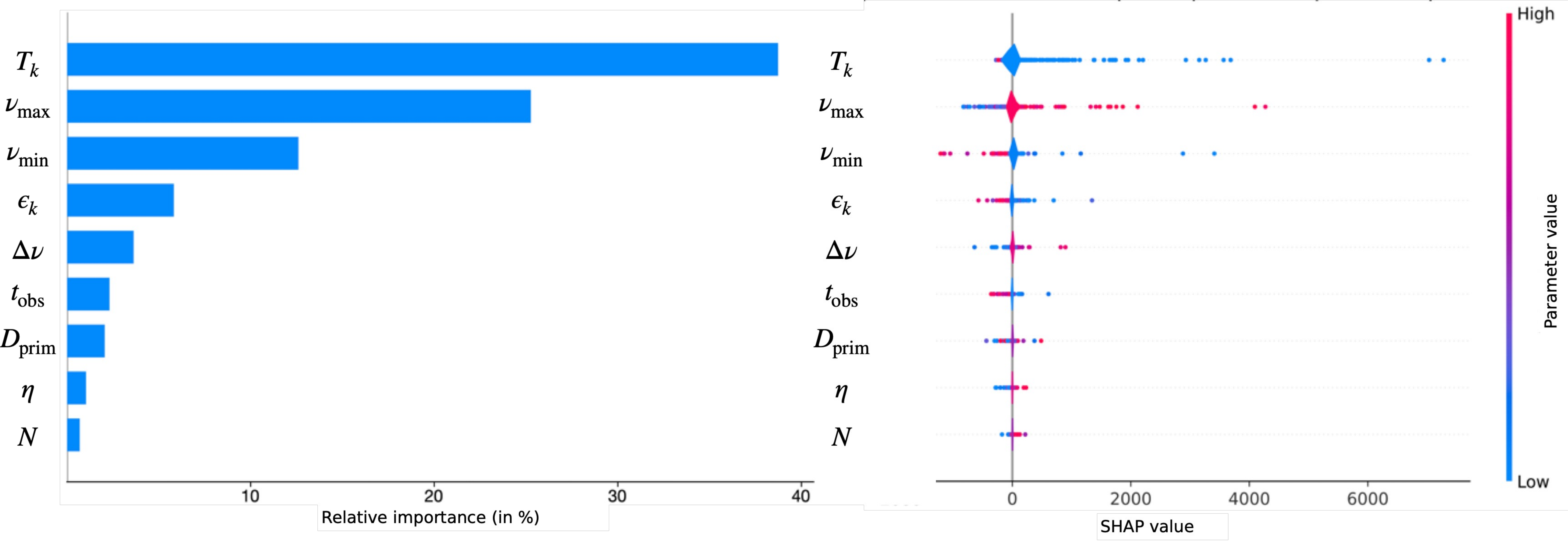}

\captionof{figure}{Same as Fig.~\ref{fig: SHAP y}, but for $A_{\mathrm{CIB}}$.}
\label{fig: SHAP cib}

\end{center}

The corresponding SHAP-based parameter importance and feature-dependence plots are shown in Figs.~\ref{fig: SHAP mu} and~\ref{fig: SHAP cib}, respectively. These results complement the $y$-distortion analysis presented in Fig.~\ref{fig: SHAP y} and provide a comparison of the instrumental parameter dependencies across the three observables.

\end{appendix}
\end{document}